\documentclass[12pt]{article}	\def\VSN{{\today~@ \printTIME ~/ \bf\jobname}}

\usepackage{latexsym,amssymb,amsmath,bbm,cite,mathrsfs,rotating,hyperref,amsfonts}   

\long\def\new#1\endnew{{\bf #1}}		\long\def\del#1\enddel{}

\def\ifundefined#1{\expandafter\ifx\csname#1\endcsname\relax}

\def\BE {\begin{equation}}      \def\EE {\end{equation}}
\def\BEA{\begin{eqnarray}}      \def\EEA{\end{eqnarray}}     \let\5=\overline
\def\BI {\begin{itemize}}       \def\EI {\end{itemize}}		\let\ex=\times

\def\HS#1 {\hspace*{#1pt}} \def\VS#1 {\vspace*{#1pt}} \long\def\del#1\enddel{}
\def\BC{\begin{center}}   \def\VR#1#2{\vrule height #1mm depth #2mm width 0pt}
\def\EC{\end{center}}      \def\TVR#1#2{@{~~\VR{#1}{#2}}}      

\def\IP{{\mathbb P}}	\def\IZ{{\mathbb Z}}		\def\id{\mathbbm1}

\def\fBM#1\fEM{\hbox{\fns$\begin{matrix}#1\end{matrix}$}}
\def\tBM#1\tEM{\hbox{\tiny$\begin{matrix}#1\end{matrix}$}}

\let\a=\alpha   \let\b=\beta    \let\g=\gamma   \let\d=\delta   
         \let\th=\theta  
\let\k=\kappa     \let\m=\mu      
\let\n=\nu            \let\p=\pi            
\let\t=\tau        \let\c=\chi      \let\ps=\psi 
\let\Ph=\phi    \let\PH=\Phi            
          \let\G=\Gamma   

\newcounter{TRefNX} \let\OLDcite=\cite  \makeatletter
\def\makeTRefs#1{\@for  \NewTRef:=#1\do{\global\makeTRef{\NewTRef}}}
\def\makeTRef#1{\ifundefined{TRef#1}\stepcounter{TRefNX}%
\expandafter\xdef\csname TRef#1\endcsname{\theTRefNX}\fi}\makeatother
\def\NEWcite#1{\makeTRefs{#1}\OLDcite{#1}}

   \newcount\HOUR  \HOUR=\the\time \divide\HOUR by 60    \multiply \HOUR by 60
   \newcount\MIN   \MIN=\the\time  \advance\MIN by -\HOUR   \divide\HOUR by 60
   \ifnum   \MIN>9  \def\printTIME{{\it\the\HOUR\,:\,\the\MIN}}
         \else   \def\printTIME{{\it\the\HOUR\,:\,0\the\MIN}} \fi 
\ifundefined{draftmode} {} \else        
\let\cite=\NEWcite 
\del
	   
\enddel
   \def\LLab#1{\BP(0,0)\unitlength=1mm\put(-9,-1.6){\makebox(0,0)[cr]{\tiny #1
        \rlap{$_{_{\makeatletter\csname TRef#1\endcsname\makeatother}}$}}}\EP}
   \def\BP{\begin{picture}} \def\EP{\end{picture}} \let\OLDbib=\bibitem 
   \def\NEWbib#1{\OLDbib{#1}\LLab{#1}} \let\bibitem=\NEWbib
\fi

\let\fns=\footnotesize            

\def\Rgs{{{\bf R_0}}}
\def\CP{{\bf c}}
\def\ACP{{\bf a}}

\def\N{\mathcal{N}}
\def\vector{\mathrm{v}}
\def\spinor{\mathrm{s}}
\def\conjspinor{\mathrm{\bar{s}}}

\def\GSO{{GSO}}
\def\Z{\mathbb{Z}}
\def\Sbar{\bar{S}}
\def\sbar{\bar{s}}
\def\hbar{\bar{h}}

\def\Qbar{\overline{Q}}
\def\qbar{\bar{q}}
\def\intern{{int}}
\def\st{{st}}

\def\LC{{LC}}
\def\C{\mathcal{C}}
\def\F{\mathcal{F}}
\def\G{\mathcal{G}}
\def\H{\mathcal{H}}
\def\M{\mathcal{M}}

\newcommand{\be}{\begin{equation}}
\newcommand{\ee}{\end{equation}}
\newcommand{\ba}{\begin{eqnarray}}
\newcommand{\ea}{\end{eqnarray}}
\newcommand{\ovl}{\overline}
\newcommand{\half}{\frac{1}{2}}

\usepackage{color}

\usepackage{color}

\usepackage{color}

\usepackage{color}

\numberwithin{equation}{section}

\begin{document}          \baselineskip=16pt

\title{\hfill {\small{\begin{flushright} TUW-10-14 \\ IPhT-T10/159 \end{flushright}}}\vskip 1cm
Counting charged massless states in the \\ (0,2) heterotic CFT/geometry connection
}

\author{Matteo Beccaria$^{a}$\footnote{matteo.beccaria AT le.infn.it}, \ Maximilian Kreuzer$^{b}$\footnote{kreuzer AT hep.itp.tuwien.ac.at} \  and Andrea Puhm$^{bcd}$\footnote{puhma AT hep.itp.tuwien.ac.at or andrea.puhm AT cea.fr}}

\date{}

\maketitle

\begin{center}
\emph{
$^{a}$
Physics Department, Salento University and INFN, 73100 Lecce, Italy \\
\vskip 0.04cm
$^{b}$ Institut f\"ur theoretische Physik, TU Vienna, \\  Wiedner Hauptstrasse 8-10, 1040 Vienna, Austria\\
\vskip 0.04cm
$^{c}$  Institut de Physique Th\'{e}orique, CEA/Saclay, \\ 91191 Gif-sur-Yvette Cedex, France\\
\vskip 0.04cm
$^{d}$Kavli Institute for Theoretical Physics, Kohn Hall, UCSB, \\Santa Barbara, CA 93106, USA
\vskip 0.04cm}
\end{center}

\begin{abstract}
We use simple current techniques and their relation to
	orbifolds with discrete torsion for studying the (0,2)
	CFT/geometry duality with non-rational internal $\N=2$ SCFTs.
	Explicit formulas for the charged spectra of heterotic $SO(10)$ GUT models 
	are computed in terms of their extended Poincar\'e polynomials
	and the {complementary Poincar\'{e}} polynomial which can be computed in terms of the elliptic genera.
	While non-BPS states contribute to the charged spectrum, their 
	contributions can be determined also for non-rational cases. 
	For model building, with generalizations to $SU(5)$ and SM gauge groups,
	one can take advantage of the large class of Landau-Ginzburg orbifold examples.
\end{abstract}

\newpage
\tableofcontents

\section{Introduction}

A beautiful example of the interplay between world-sheet and space-time 
techniques is the Greene-Plesser (GP) mirror construction 
\cite{Greene:1990ud}, which identifies charge conjugation for (tensor
products 					
of) minimal 
models 	
with an orbifold 
construction and thus establishes the mirror automorphism for an exactlty 
solvable point in the moduli space of a string compactification.
Deformation arguments can then be used to extend mirror symmetry to the 
geometrical realm. 

The setting of the GP construction is a heterotic string whose
compactification  geometry is replaced by a tensor product 
$\mathcal C_{int}=C_{k_1}\otimes\ldots \otimes C_{k_r}$ of $\N=2$ 
superconformal minimal models with central charge $c=9$, for which Gepner 
\cite{Gepner} was able to construct a modular invariant partition function 
with space-time
supersymmetric massless particle spectrum and gauge group $E_6\times E_8$.
The relation to geometry proceeds via the Landau--Ginzburg (LG) description
\cite{Greene:1988ut,Lerche:1989uy} of minimal models by Fermat-type
superpotentials $W=\PH_1^{K_1}+\ldots+\PH_r^{K_r}$ with $K_i=k_i+2$, which
can then be identified with the hypersurface equation $W=0$ defining
a Calabi--Yau variety in a weighted projected space. 
More precisely,	the exactly solvable {\it Gepner point} is located
at small values of the K\"ahler moduli 
and can be reached as a certain limit in the parameter space of the 
gauged linear sigma model (GLSM) \cite{Witten:1993yc}.

Mirror symmetry has been a pivotal tool in the study of 
non-perturbative physics for two decades and is well understood for
heterotic (2,2) compactifications 
\cite{Candelas:1990rm,Batyrev:1994hm,Batyrev:1995ca}. 
From the phenomenological
perspective, however, (0,2) world-sheet supersymmetry (with quantized
charges) is sufficient for low energy space-time SUSY and much more
attractive models with realistic GUT gauge groups arising quite naturally.
The GLSM provided an important step for the construction of such models
as it allowed the study of (0,2) deformations away
from the (2,2) locus \cite{Witten:1993yc} as well as the construction of 
large classes of genuine (0,2) models with geometrical and
Landau-Ginzburg phases, like the Distler-Kachru (DK) models \cite{di94}.

On the rational CFT side a powerful formalism generalizing
Gepner's construction was developed by Schellekens and 
Yankielowicz  \cite{sc90} in terms of simple currents \cite{sc90r}, 
which are related to certain discrete symmetries and, in a sense, 
can be regarded as a generalization of free fields. String vacua, from
this perspective, are constructed by starting with tensor products of 
CFTs and performing a number of projections, like generalized GSO
or alignments of Ramond sectors. All of these projections can be realized
as simple current modular invariants (SCMIs) of extension type 
\cite{sc90,kr95,wkm} and large classes of (0,2) models can be constructed 
very naturally with the same techniques. Moreover, the general classification
of SCMIs \cite{ks93} uncovered their relation to orbifolds with discrete 
torsion, enabling translations of results into geometrical language and
suggesting generalizations beyond the rational realm \cite{kr95}.

Like in the case of (2,2) models, a comparison of particle spectra
can be performed to look for identifications of models that
are constructed with geometry and CFT methods, respectively.
In \cite{bl95} Blumenhagen and Wi{\ss}kirchen (BW) indeed discovered a (0,2)
cousin of the quintic with 80 generations and gauge group SO(10) that
showed up on either side, and the construction could be extended to
a whole family of identifications \cite{bl96,bsw96,kr96}. On the CFT
side it is based on a Gepner-type tensor product, but with an additional
simple current $J_b$ that acts as a $\IZ_2$ twist breaking the $E_6$ 
gauge group of the (2,2) model down to $E_5\cong SO(10)$. On the 
geometry side this corresponds to a rank 4 vector bundle $E$ on a Calabi-Yau 
manifold $X$ whose data are constrained by the anomaly matching condition 
$c_2(E)=c_2(X)$ and make sense also for certain non-rational theories
$\mathcal C_{int}$ like Landau-Ginzburg models and orbifolds thereof.
More precisely, there is a conjectured identification between certain
rational (0,2) heterotic strings constructed with simple current
techniques and (0,2) Landau-Ginzburg models, which can then be deformed
to large volume in terms of their GLSM realization. The latter is an
interesting topic on its own but is beyond the scope of this note.

In the present note we investigate the non-rational generalization of the
CFT/geometry connection proposed in \cite{kr96,wkm} and
develop  tools for the computation of their massless spectra on 
the CFT side. Our starting point is the identification of simple current 
modular invariants with orbifolds with discrete torsion \cite{ks93,wkm}, 
which can be used to reformulate the construction of Blumenhagen et al.
\cite{bl95,bl96} in a more geometrical language and to extend it, 
for example, to arbitrary internal $\N=2$ SCFTs containing a minimal 
model factor at odd level. The breaking of $E_6$ 
to the gauge group $D_5=SO(10)$ by a simple current $J_b$ is thus attributed 
to discrete torsions spoiling the algebra extension in the gauge sector
and corresponds to a $\IZ_2$ orbifolding. The main technical point
will be the computation of the spectrum in $J_b$-twisted sectors, for which
non-BPS states turn out to contribute even to non-gauge-singlet massless
states.

Our construction also has interesting implications for (0,2) mirror symmetry
\cite{bsw96,bsf,Sharpe:2008rd,Kreuzer:2010ph,Melnikov:2010sa} because
charge conjugation is a simple current modular invariant
for (tensor products of) $\N=2$ minimal models.
According to the general classification \cite{ks93},
the data defining a SCMI is a simple current (or orbifold) group 
$\mathcal G$ together with a choice of discrete torsions (in terms of a
fractionally quantized matrix $X_{\mathcal G}$ with given symmetrization). 
Since products of SCMIs are again SCMIs the mirrors of our (0,2) models 
can be explicitly constructed within the same framework, which should explain 
the large degree of mirror symmetry for orbifold spectra observed in 
\cite{bsw96,bsf}. 
By our extension of the formalism to non-rational models
this (0,2) version of the Greene-Plesser construction extends to the 
Berglund-H\"ubsch mirror construction for Landau-Ginzburg orbifolds with
minimal transversal superpotentials \cite{Berglund:1991pp}. The precise
mirror map for untwisted minimal LG models has been constructed in
\cite{Kreuzer:1994uc} and can be extended to arbitrary orbifolds with discrete
torsion using the methods developed in \cite{odt} by relating discrete
torsion to the modding of quantum symmetries \cite{va891}. This generalizes
and must be consistent with the SCMI mirror construction, but in both
versions only an algorithm but no explicit formulas for the twist groups
and torsions of the mirror are available. The universality of these
constructions suggests, however, that a purely group-theoretical description
should exist and would be very interesting to be unveiled.

In section 2 we define our class of models and recollect the basis of our 
formalism, which at the same time generalizes and simplifies Gepner's 
construction within RCFT, and embeds it beyond rationality to orbifolding 
techniques via the classification of SCMIs.
In section 3 we work out explicit formulas for non-singlet matter spectra. In our
class of models the breaking of (2,2) to (0,2) models with GUT gauge
is due to a twist that spoils alignment of Ramond and Neveu-Schwarz sectors 
for the left-movers. As a consequence, it turns out that non-BPS states
contribute even to charged matter. 
Using the simple current orbit structure
and spectral flow we can determine, however, everything in terms of 
the finite data given by charge degeneracies of Ramond ground states and excited Ramond states of an
arbitrary $\N=2$ SCFT, as encoded in its extended Poincar\'{e} polynomial (EPP) and the {complementary Poincar\'{e}} polynomial (CPP).
In section 4 we discuss the geometry connection and check the
correspondence of spectra for non-rational examples. Examples and some details of the construction are collected in section 5 and the appendices.

\def\mao#1{\mathop{\rm #1}\nolimits}	\def\rel#1 #2{\buildrel #1 \over {#2}}
\def\2{{\textstyle\frac12}}		\let\6=\partial	\let\then=\Rightarrow
	
\def\ca{{\mathcal A}}	\def\cc{{\mathcal C}}	\def\cg{{\mathcal G}}	
\def\co{{\mathcal O}}	\def\cu{{\mathcal U}}	\let\ket=\rangle

\section{Simple currents, orbifolds, and (0,2) models}

In this section we recollect the ingredients of our construction, as reviewed
in more detail in \cite{wkm}.
The discussion is intended to provide an intuitive picture rather than
proofs, which can be found in the references.
We start with simple currents and
their relations to orbifolds and then discuss their application to projections
in arbitrary $\N=2$ SCFT, with a summary of what we need for the special case
of minimal models. Then we define our class of (0,2) models and
discuss space-time SUSY ({\it i.e.} the generalized GSO projection) and the
breaking of the gauge group by a simple current $J_b$, which we will
refer to as the ``Bonn twist''.

\subsection{Simple currents and orbifolds with discrete torsion}

The left-{\it chiral algebra} (or {\it vertex algebra}) $\mathcal A_L$ of a 
conformal field theory is the holomorphic subalgebra of the operator algebra.
Similarly, the anti-holomorphic fields define the right-chiral algebra $\mathcal A_R$.
The Hilbert space of states $\mathcal H$ can thus be organized into 
representations of the symmetry algebra $\mathcal A_L\otimes A_R$, with chiral and 
antichiral labels $a$ and $\bar b$, respectively, labeling characters 
$\c_a(\t)=\text{tr}_{\H_a}\exp\left(2\p i\t (L_0-\frac c{24})\right)$ and
their right-moving partners $\c_{\bar b}(\bar\t)$.
If the decomposition 
$\mathcal H=\bigoplus_{a,\bar b} {\H}_a\otimes {\H}_{\bar b}$ is finite the 
conformal field theory is called rational and the 1-loop partition function
$Z(\t)=M_{a\bar b}\c_a(\t)\c_{\bar b}(\bar\t)$ can be written in terms
of a finite non-negative integer matrix $M_{a\bar b}$ of multiplicities,
called modular invariant,%
\footnote{~ 
	Modular invariance, in this context, usually refers to the conditions
	$[M,T]=[M,S]=0$ for the representation matrices $T$ and $S$ of the 
	respective $SL(2,\IZ)$ generators on the characters. The full 
	consistency conditions of conformal field theories require, in 
	addition, appropriate behavior of all correlation functions under 
	factorization and mapping class group transformation of Riemann
	surfaces of arbitrary genus, which fortunately can be shown to follow
	from a finite number of constraints (like 2-loop modular invariance
	or modularity of 1-point functions on the torus).
}
with a unique identity 
$M_{\id\id}=1$.

It will be important below to distinguish between individual {\it conformal
fields}, labeled by their full set of quantum numbers, and {\it conformal
families} $\Ph_{a\bar b}(z,\bar z)$, which consist of all conformal fields 
corresponding to a representation ${\H}_a\otimes {\H}_{\bar b}$. For simplicity
we can think of the diagonal modular invariant as our starting point
and only consider left-moving labels $a$ (or, more rigorously, ignore
the ``chiral'' $\Ph_{a}$ altogether and only refer to representation 
labels $a$). 
From the operator product algebra we can then extract the associative and 
commutative fusion algebra 
$\Ph_a\times \Ph_b=N_{ab}{}^c\Ph_c$, whose non-negative integer structure 
constants $N_{ab}{}^c$ denote the multiplicity of the field  $\Ph_c$
in the OPE $\Ph_a\times \Ph_b$.%
\footnote{~
	This multiplicity is usually $N_{ab}{}^c\in\{0,1\}$, except if the 
	conformal Ward identities 
	do not fix all coefficients 
	of higher descendents in terms of the coefficient of the 
	most singular contribution of the family $\Ph_k$ to the OPE of
	two operators $\hat\Ph_a\in\Ph_a$ and  $\hat\Ph_b\in\Ph_b$.
	$N_{ab}{}^d C_{dc}$ is the number of independent 
	3-point conformal blocks in $\langle\Ph_a\Ph_b\Ph_c\rangle$, where 
	the charge conjugation matrix $C_{ab}$ is a symmetric permutation 
	matrix related to the fusion coefficients by $C_{ab}=N_{ab}{}^\id$. 
}
Simple currents are conformal families $J$ with a unique fusion
product, {\it i.e.} for which $J\ex \Ph_a=\Ph_{Ja}$
for a unique family $\Ph_{Ja}$ \cite{sc90}. Examples are free 
fermions or vertex operators of free bosons, so that simple currents can be 
regarded as a generalization of free fields. They decompose the set of 
conformal families into orbits which are of finite length
\BE
	\Ph_a\to \Ph_{Ja}\to \Ph_{J^2a}\to\ldots\to \Ph_a,
\EE 
in a rational CFT. The maximal orbit length $N_J$, called the order of $J$, 
is the length of the orbit of the identity because $J^{N_J}\id=\id$ implies
that every other orbit length is a divisor of $N_J$.

Since all members of a conformal family have the same conformal weight modulo
1, uniqueness of the fusion product of $J$ implies that all branch cuts 
originating (with slight abuse of notation) from
OPE singularities of the form $(z-w)^{h_{Ja}-h_J-h_a}$ 
have the same monodromy phase $e^{-2\p i Q_J(\Ph_a)}$ about 
the singular point, where
\BE
        Q_J(\Ph_a)\equiv h_J+h_a-h_{Ja}~~~\mod~~1	\label{QJ}
\EE
is called the monodromy charge $Q_J$ of $\Ph_a$. The
important observation is that $Q_J$ is conserved modulo 1 in operator products
and thus implies the existence of a phase symmetry 
$\Ph_a\to e^{-2\p i Q_J(\Ph_a)}\Ph_a$, which is a cyclic group $\IZ_{N_J}$
of order $N_J$ because it can be shown that the charges $Q_J$ are quantized 
in units of $1/N_J$ \cite{sc90r}.

The set of all simple currents of a rational CFT forms a finite abelian 
group under fusion, called the center. In order to implement the necessary
projections for the construction of our models we will work with a fixed
subgroup $\cg$ of the center, for which we can introduce a set 
of generators $\cg=\langle J_i\rangle$ of order $N_i=N_{J_i}$. Each
current $J=\prod_i (J_i)^{\a^i}\in\cg$ can then be written as 
$J=\sum_i \a^iJ_i$ in an additive notation, where we identify 
$J\cong[\a]$ with an integer vector $\vec \a$
whose components $\a^i$ are defined modulo $N_i$. It can then be shown that
all conformal weights and monodromy charges modulo 1 of all
simple currents in $\mathcal G$ can be parametrized in terms
of a matrix $R_{ij}$ \cite{ga91},                              	\VS-3
\BE            \textstyle                                   	\VS-3
     R_{ij}=\frac{r_{ij}}{N_i}\equiv Q_i(J_j) = Q_j(J_i),\qquad
     h_{[\a]}\equiv \2\sum_i r_{ii}\a^i-\2\sum_{ij}\a^iR_{ij}\a^j   \label{Rh}
\EE
with $Q_i\equiv Q_{J_i}$ and $r_{ij}\in\IZ$. The definitions of $Q_i$ and
$R_{ij}$, in turn, imply					\VS-3
\BE                                           		  	\VS-3
     h_{[\a] a}\equiv h_{a}+h_{[\a]}-\a^iQ_i(a), \qquad
     Q_i([\a] a)\equiv Q_i(a)+R_{ij}\a^j.			  \label{hQ}
\EE
If $N_i$ is odd we can always choose $r_{ii}$ to be even. With this 
convention all diagonal elements $R_{ii}$ are defined modulo 2 for both, 
even and odd $N_i$.

A simple current modular invariant (SCMI) is a modular invariant with
$M_{ab}\neq0$ only if $b$ is on a simple current orbit of $a$, {\it i.e.} if
there is a simple current $J$ with $b=Ja$.
Because of (\ref{Rh}) and (\ref{hQ}) T-invariance $[M,T]=0$ which requires
$h_a-h_{[\a]a}\in\IZ$, implies that simple currents $J_i$ of even
order can only contribute SCMIs if $r_{ii}\in2\IZ$.
Subgroups $\cg$ of the center violating this condition can thus be excluded 
from further consideration so that 
$r_{ii}\in2\IZ$ and $h_{[\a]}\equiv-\frac12 \a^iR_{ij}\a^j$. It
can now be shown that the most general SCMI is of the form%
\footnote{~
	The proof in \cite{ks93} uses factorization and regularity
	assumptions that exclude unphysical solutions. A state-of-the-art 
	approach is based on modular tensor categories \cite{Fuchs:2002cm};
	cf. section 4.2 of \cite{Fuchs:2009iz} and references therein.
}        
\BE 	
		\boxed{\textstyle\phantom{\!\Bigl(}		\label{X}
	M_{a,[\a]a}=\m(a)\prod_i\d_\IZ\left(Q_i(a)+X_{ij}\a^j\right),~}
\EE
where $X$ is defined modulo 1 and quantized by $\gcd(N_i,N_j)X_{ij}\in\IZ$.
The multiplicity $\m(a)$ is the order of the stabilizer of $\cg$ on the
orbit of $\Ph_a$ and $\d_\IZ$ is one on integers and 0 otherwise.

The formula (\ref{X}) lends itself to an instructive and useful orbifold 
interpretation \cite{wkm},
where $\d_\IZ(Q_i+\ldots)$ is identified as the projection to states that
are invariant under the $\IZ_{N_i}$ phase symmetries implied by $J_i$ and
$\vec \a$ labels the twisted sectors. 
A simple calculation shows that level matching $h_a-h_{[\a]a}\in\IZ$ 
fixes the symmetric part $X+X^T\equiv R$ modulo 1 for off-diagonal and 
modulo 2 for diagonal matrix elements, while the 
ambiguity due to the choice of a properly quantized antisymmetric part 
of $X$ exactly corresponds to the freedom due to the choice of
discrete torsions%
\footnote{~
	Discrete torsions can be interpreted as phase ambiguities of the
	orbifold group action on twisted vacua, which are
	proportional to $\a^j$ because of the twist selection rules (also
	known as {\it quantum symmetries} \cite{va891}). In fact,
	the formula (\ref{X}) was motivated by universalities observed
	in the classification efforts of \cite{ga91} and the observation that
	proper account of quantum symmetries was vital for understanding the
	relation between orbifolds and modular invariants in Gepner models 
	\cite{Fuchs:1992fv}.
}
of the orbifolding procedure.

In conclusion we note that orbit positions $\a^i$ in SCMIs (\ref X)
generalize the shift vectors of Gepner's construction and, 
via their identification with the labels of twisted sectors, 
embed it into the framework of orbifolds, which we will use to generalize 
heterotic (0,2) models to the non-rational realm on the CFT side of
the proposed geometry/CFT duality.

\subsection{Universal currents in $\N=2$ superconformal field theories}

In non-geometrical supersymmetric compactifications the sigma-model on 
a Calabi-Yau is replaced by an ``internal'' $\N=2$ SCFT $C_{int}$ with $c=9$
and a number of projections like charge quantization (or generalized GSO)
and the alignment of spinors with the Ramond sector, which we will
discuss in turn.
The $\N=2$ algebra is generated by the Fourier modes of the
energy momentum tensor $T(z)$, its fermionic superpartners $G^\pm(z)$, and 
a $U(1)$ current $J(z)$. For unitary theories positivity of
expectation values of the anticommutator
\BE	\textstyle
	\{G_r^-,G_s^+\}=2L_{r+s}-(r-s)J_{r+s}+\frac c3(r^2-\frac14)\d_{r+s},
\EE
of the superconformal charges implies the inequalities \cite{Lerche:1989uy}
\BE
	h_{R}\ge c/24,\quad h_{NS}\ge |q/2| \qquad\text{with}\qquad\label{BPS}
	L_0|h,q\rangle=h|h,q\rangle, \quad J_0|h,q\rangle=q|h,q\rangle
\EE
for $r=s=0$ in the Ramond sector $r,s\in\IZ$ and for $r=-s=\pm1/2$ in 
Neveu-Schwarz sector $r,s\in\frac12+\IZ$, respectively. These inequalities 
are saturated by the ``BPS states'' 
\BE	\textstyle
	|\Rgs \rangle=|h=\frac{c}{24},q\rangle,			\label{RCP}
\qquad  |\CP \rangle=|h,q=2h\rangle,
\qquad  |\ACP \rangle=|h,q=-2h\rangle,
\EE
called Ramond ground states and (anti)chiral primary states and are defined by
$G_0|\Rgs\ket=0$, $G_{-1/2}^+|\CP\rangle=0$ and 
$G_{-1/2}^-|\ACP\rangle=0$, respectively (in addition to being primary!).
For (2,2) heterotic strings, 
these states completely determine the {\it charged massless} spectrum.

The $\N=2$ algebra admits a continuous family of automorphisms known as
spectral flow,
\BE \textstyle\hbox{\small$
        L_n\rel {\cu_\th} \longrightarrow       
        L_n+\th J_n+\frac c6\th^2\d_n,~~~~~
    J_n\rel {\cu_\th} \longrightarrow   
        J_n+\frac c3\th\d_n,~~~~~
        G_r\rel {\cu_\th} \longrightarrow
                G_{r\pm\th}^\pm$}, 
\EE
which interpolates between the Ramond and the NS sector. In particular,
$\cu_{\pm1/2}$ maps Ramond ground states to chiral and antichiral
primary fields, respectively. Spectral flow is best understood by 
bosonization of the $U(1)$ current $J(z)=i\sqrt{\frac c3}\,\6X(z)$ in terms 
of a free field $X$ with normalization $J(z)J(w)\sim \frac c3/(z-w)^2$. 
A charged operator $\co_q$ can thus be written as a 
normal ordered product of a
vertex operator with a neutral operator $\co_0$,                \vspace{-3pt}
\BE                                                                     \VS-3
        \co_q=e^{i\sqrt{\frac3c}\,qX}\,\co_0(\6X,\ldots,\ps,\ldots)
\EE
with the $U(1)$ charge corresponding to the momentum of the vertex operator,
whose contribution to $h$ is $\frac{3q^2}{2c}$. The inequalities (\ref{BPS}) 
hence imply that the maximal charges of $\Rgs$ and $\CP$ states in unitary 
theories are $c/6$ and $c/3$, respectively. 

We now have all ingredients to discuss the universal center of $\N=2$ SCFT's
\cite{kr95}.
Already for $\N=1$ the supercurrent $G$ is a simple 
current, which we denote by $J_v$. Its monodromy charge is $Q_v=0$ 
for NS fields and $Q_v=1/2$ for Ramond fields since $h_v=3/2$ and the 
conformal weights of superpartners differ by integers in the Ramond sector 
and by half-integers for NS states. For $\N=2$, in addition, the Ramond 
ground state $J_s=e^{i\sqrt{c/12}\,X}$ with maximal charge $c/6$ is a 
pure vertex operator and hence a simple current. 
A short calculation shows that its monodromy charge 
is $Q_s\equiv-\frac12q$ modulo 1. If the $U(1)$ charges $q$ are quantized in 
units of $1/M$ in the NS sector then $c=3k/M$ for some integer $k$. Since
the $U(1)$ charges are shifted by $-c/6=-k/2M$ in the Ramond sector,
the order $N_s$ of $J_s$ is $2M$ if $k\in2\IZ$ and $4M$ if $k\not\in2\IZ$
and the relation between $Q_s$ and $Q_v$ modulo 1 implies
\BE	\textstyle						\label{N=2GC}
	J_s^{2M}=J_v^k,\quad J_v^2=\id \qquad\text{with}\qquad c=3k/M,
					\quad\langle J_s,J_v\rangle\cong
	\begin{cases}~~~\IZ_{4M} & \text{for }~ k\not\in 2\IZ\\
		\IZ_{2M}\ex\IZ_{2} & \text{for }~ k\in 2\IZ\end{cases}
\EE
so that the order of the universal center is $4M$ in both cases.
The best way to compute the monodromy matrix 
\BE
	R_{J_vJ_v}=1,\quad
	R_{J_vJ_s}=1/2,\quad R_{J_sJ_s}=-c/12
		+\begin{cases}0&k\in4\IZ\\1&k\not\in4\IZ\end{cases}
\EE
is by first evaluating $Q_i(J_j)$ modulo 1 using
$h_{J_v}=3/2$, $h_{J_s}=c/24$, $h_{J_s^2}=c/6$ and then fixing the diagonal
$R_{ii}$ modulo 2 by imposing $h_{J_i}\equiv-\frac12 R_{ii}$ for 
$r_{ii}\in2\IZ$.

\subsubsection{Minimal models, field identifications and mirror symmetry}
\label{sec:MM}

The chiral labels $a=(l,m,s)$ of $\Ph_a\equiv\Ph_m^{ls}$ for minimal models 
$\cc_k$ at level $k$ are best understood from their coset representation
\BE                                                             \VS-5
        \cc_k=(SU(2)_k\ex U(1)_2)/U(1)_{K}\quad~~\text{with}\quad K=k+2
	\quad \text{and}\quad c=3k/K. 
\EE
The labels $l=0,\ldots,k$ and $s$ mod 4 refer to the factors 
$SU(2)_k\ex U(1)_2$ in the numerator, and the $U(1)_{K}$ label $m$ is 
defined modulo $2K$ in accord with the convention that $U(1)_{K}$ has $2K$
representations.
Ramond and NS fields correspond to odd and even $s$, respectively.
The conformal weights and the $U(1)$ charges obey      
\BE     \textstyle                                                   
        h\equiv\frac{l(l+2)-m^2}{4K}+\frac{s^2}{8}~\mod~1,
		\qquad\quad
        q\equiv\frac{s}{2}-\frac{m}{K}~\mod~2                \label{hqMM}
\EE 
with exact equality in the standard range $|m-s|\le l$, $-1\le s\le 1$
\cite{sc90,wkm}. 

The fusion rules of $U(1)$ and $SU(2)_k$ imply that
$\Ph_m^{ls}$ is a simple current if $l=0$ or $l=k$.
The branching rule $l+m+s\in2\IZ$ of the coset implies the necessity
of field identifications
\BE             \label{fieldID}                  \VS-3
        \Ph_m^{ls}\sim\Ph_{m+K}^{k-l,s+2}=J_{id}\ex\Ph_m^{ls} 
	\qquad\hbox{with}\qquad 
                J_{id}=\Ph_K^{k2} ~~~~~\Rightarrow~~~~~ Q_{id}\equiv (l+m+s)/2
\EE
which can again be understood as a SCMI because integral monodromy 
$Q_{id}\in\IZ$ of the identification current $J_{id}$ provides the 
correct selection rule and, since $h_{id}\in\IZ$, extends the chiral 
algebra \cite{sc90r}.

After field identification we find that the center of $\cc_k$ is exactly
the generic center of an $\N=2$ SCFT with
\BE
	J_s:=\Ph_1^{01}\sim\Ph_{1-K}^{k3},\qquad 
	J_v:=\Ph_0^{02}\sim\Ph_K^{k0}\qquad \text{and}\qquad M=k+2=K. 
\EE
Note that the general parametrization $c=3k/M$ of the central charge was
chosen above in order to emphasize the analogy of $k$ with the level of the 
minimal model, namely that $J_s^{2M}=J_v^k$ determines the group structure 
(\ref{N=2GC})
of the center, while the inverse charge quantum $1/M$ is in general 
unrelated to $k$.

The Landau-Ginzburg description of a minimal model $\cc_k$ requires a 
simple chiral superfield $\PH$ with superpotential $W=\PH^K$ whose
chiral ring \cite{Lerche:1989uy} is generated by $\PH$ modulo 
$\6W\sim \PH^{K-1}$. We hence expect $k+1$ chiral primary fields $\PH^l$, 
whose chiral labels are easily checked to be $\Ph_{-l}^{l,0}$ by comparing 
charges, conformal weights and fusion rules. The remaining BPS states
\BE	\hbox{\small\!\!\!\!
	\begin{tabular}{c|c|c}			
\VR{4}{2.5}	\!\!anti-chiral primary:\; $q=-\frac lK$ &  
		R ground states:\; $q=\pm(\frac{c}{6}-\frac{l}{K})$ &
		chiral primary:\; $q=\frac lK$\!\!		\\ \hline
\VR51   $\Ph_{l}^{l0}\sim\Ph_{K+l}^{k-l,2} \sim \5\PH^l $& 
   $\Ph_{\pm(l+1)}^{l,\pm1}\sim\Ph_{\mp(k-l+1)}^{k-l,\mp1}$&
        $\Ph_{-l}^{l0}\sim\Ph_{K-l}^{k-l,2}\sim \PH^l$	\\
	\end{tabular}}
\EE
can then be identified, for example, by charge conjugation and spectral flow.

It is instructive to study the orbit structure of the center for minimal
models. Taking into account the selection rule $Q_{id}\in\IZ$ and
field identifications we have $2K(k+1)$ chiral labels and $4K$ simple 
currents so that we have to expect fixed points for $k\in2\IZ$.
Indeed, since $J_s^\n J_v^\a\ex \Ph_m^{l,s}=\Ph_{m+\n}^{l,s+\n+2\a}$
the orbits are parametrized by $l$, which can be restricted to $l\le k/2$
because field identifications map $l\to k-l$, which leads to an orbit 
of length $2K$ with multiplicity $\m(l)=2$ stabilized by 
$\Ph_K^{02}=J_s^KJ_v^{k/2}$ for $l=k/2$ if $k\in2\IZ$.\\

Note that in general each orbit contains 
exactly two BPS states of each type. Considering, for example, chiral primaries $\Ph_{-l}^{l0}$ we use field identification to find its partner with $l'=k-l$ at the orbit position
$\Ph_{-l'}^{l'0}=J_s^{2(l+1)}J_v^l\Ph_{-l}^{l0}$. 
For mirror symmetry we, instead, need to implement charge conjugation 
$\Ph^{ls}_m \to \Ph^{l,-s}_{-m}$ by fusion with a simple current $\Ph_{-2m}^{0,-2s}=J_s^{-2m}J_v^{m-s}$ with $m \mod K$ and $s \mod 4$. Due to the orbit structure charge conjugation is a SCMI, denoted by 
$C_{a,Ja}$, which is determined by the group $\G$ and the discrete torsion $X$ of the orbifolding procedure. A convenient choice of basis for the generators of the group is $\G=\langle J_1=J_s^2 J_v=\phi^{0,0}_{2}, J_2=J_v=\phi^{0,2}_{0} \rangle$ because the SCMI then splits according to $C_{a,Ja}=C_{m\to -m} \times C_{s\to -s}$. From (\ref{SCMI}) we can calculate the symmetric part of the torsion matrix $X_{(ij)}=\frac{R_{ij}}{2}$ to be $X_{11}=\frac{1}{K}$ and $X_{22}=-\frac{1}{2}$, while the antisymmetric part corresponding to the discrete torsion in the orbifolding procedure vanishes.

\subsection{Symmetries and projections for  $(0,2)$ heterotic models}

Let us review the structure of a generic four-dimensional compactification of the (2,2) heterotic string.
The right-moving sector consists of four space-time coordinates and their superpartners $(X^{\mu},\5\psi^{\mu})$, a ghost plus superghost system $(b,c,\b,\g)$, and an "internal" $\N=2$, $c=9$ SCFT $\C_\intern$ which is the abstract version of a supersymmetric sigma model on a Calabi-Yau. The left-moving sector is a bosonic string with space-time plus ghost part $(X^{\mu},b,c)$ and the same internal sector $\C_\intern$ so that a left-moving CFT with central charge 13 needs to be added for criticality. Modular invariance requires this CFT to be either an 
$\hat{E}_8 \times \hat{D}_5$ or $\hat{D}_{13}$ level 1 affine Lie algebra, where we will henceforth ignore the phenomenologically less attractive $\hat{D}_{13}$. Instead of this covariant quantization we can also use light-cone gauge, which amounts to ignoring the (super-) ghosts and restricting the space-time coordinates to transverse directions.
We thus have two components $(\mu=2,3)$ of the space time bosons $X^{\mu}(z, \5z)$ and fermions
$\5\psi^{\mu}(\5z)$. The right-moving sector is a conformal field theory with $\5c=12$ composed by 

{
\renewcommand{\arraystretch}{1.5}
\renewcommand{\tabcolsep}{0.2cm}
\begin{tabular}{lll}
two copies of the free right-moving  SCFT $(X, \5\psi)$ &:& $\5c$ = $2\times\frac{3}{2}$ = 3, \\
an internal ${\cal N}=2$ SCFT with the central charge &:& $\5c$ = 9.
\end{tabular}
}

\noindent
The left-moving sector is a conformal field theory with $c=24$ composed by 

{
\renewcommand{\arraystretch}{1.5}
\renewcommand{\tabcolsep}{0.2cm}
\begin{tabular}{lll}
two copies of the free left-moving boson CFT &:& c = 2, \\ 
an $(\widehat E_{8})_{1}\times (\widehat D_{5})_{1}$ Ka\v{c}-Moody algebra
&:& c = 8+5 = 13, \\
an internal ${\cal N}=2$ SCFT with the central charge &:& c = 9.
\end{tabular}
}

\noindent
In the context of a sigma model on a Calabi-Yau manifold the superstring vacuum is then obtained by aligning space-time spinors and tensors with internal Ramond and Neveu-Schwarz sectors, respectively, and carrying out the (generalized) GSO projection. This can be understood in terms of SCMI's of extension type which we will discuss below.

In order to apply simple current techniques \cite{sc90}, as introduced in the previous sections, to our heterotic $(0,2)$ models, we start with a left-right symmetric theory which can be achieved by applying the so-called 
Gepner map to the right-movers.
This map dates back to \cite{cent85,ens86}. The fact that it preserves modular invariance and spin-statistics
signs in the partition function was proved in the context of the covariant lattice construction \cite{lls87}.
Later, it was applied by Gepner in order to relate type-II superstrings to heterotic strings \cite{Gepner}.

 Using the language of simple currents and SCMI's we will then be able to carry out the (generalized) GSO projection and break the gauge group $E_6$ of Gepner's construction \cite{Gepner} to $SO(10)$ by the means of a simple current $J_b$, which we call the \textit{Bonn twist}. World-sheet supersymmetry will be accordingly reduced from $(2,2)$ to 
$(0,2)$.

\subsubsection{Gepner map and  generalized GSO projection in (2,2) models}

The right-moving free space-time 
fermions form a representation of $(\widehat{D}_{1})_{1}$. 
The spectrum falls into representations of this algebra which must be unbroken being the light-cone gauge
remnant of Lorentz invariance. The one loop partition function is a product of the contributions from 
the space time fields (bosons and fermions), the internal SCFT and the left-moving $(\widehat E_{8})_{1}\times (\widehat D_{5})_{1}$ Ka\v{c}-Moody algebra. For application of SCMI techniques it is convenient to 
cast the theory in a left-right symmetric form. The asymmetry is focused on the following factors
$$
\mbox{left-movers} : (\widehat E_{8})_{1}\times (\widehat D_{5})_{1}, \qquad
\mbox{right-movers} : (\widehat D_{1})_{1}\ \ (\mbox{from}\  \5\psi(\5z))
$$
Symmetry can be achieved by exploitng a remarkable map that exchanges space time fermions
with compactified internal bosons while preserving modular invariance \cite{Gepner}.
Thus, it can map a fully bosonic partition function to a superstring or heterotic one.
Conversely, starting from a heterotic partition function, we can apply the map to the right-moving 
sector and obtain a left-right symmetric theory suitable for simple current techniques \cite{sc90}.

The affine algebra {$(\widehat{D}_{n})_{1}$} has four integrable highest weight representations, the singlet $\id$, the vector $\vector$, the spinor $\spinor$ and conjugate spinor $\conjspinor$. The only integrable representation of $(\widehat E_{8})_{1}$ is the singlet $\id$.
The heterotic one loop partition function involves a bilinear
combination of the {$(\widehat{D}_{n})_{1}$} characters of  representations with $n=5$ from the left-movers
and $n=1$ from the right-movers. We can arrange the characters in a vector 
\mbox{{\boldmath $\chi$} $= (\id, \vector, \spinor, \conjspinor)$}. Let us look at the modular transformation properties of {\boldmath$\chi$}.
Under $S:\tau\to -\frac{1}{\tau}$, we have 
\be
\mbox{\boldmath$\chi$}(-\frac{1}{\tau}) = S_{2n}\,\mbox{\boldmath$\chi$}(\tau), \qquad S_{2n}
= \frac{1}{2}\,\left(\begin{array}{cccc}
1 & 1 & 1 & 1 \\
1 & 1 & -1 & -1 \\
1 & -1 & i^{-n} & -i^{-n} \\
1 & -1 & -i^{-n} & i^{-n} 
\end{array}\right).
\ee
Under $T:\tau\to \tau+1$, we have
\be
\mbox{\boldmath$\chi$}(\tau+1) = T_{2n}\,\mbox{\boldmath$\chi$}(\tau), \qquad T_{2n} = 
e^{-i\,\pi\,n/12}\,\mbox{diag}(1, -1, e^{i\,\pi\,n/4}, e^{i\,\pi\,n/4}).
\ee
The singlet of $(\widehat E_{8})_{1}$ is invariant under $S$ and gets the phase $e^{-2\pi\,i/3}$ under 
$T$~\footnote{In general, under $T$, the affine character of $\widehat{\mathfrak{g}}_{k}$ associated to the
integrable weight $\widehat\lambda$ gets the phase $e^{2\pi\,i\,m_{\widehat{\lambda}}}$ where the modular
anomaly $m_{\widehat{\lambda}}$ can be expressed in terms of the Weyl vector $\rho$ and dual Coxeter number $g$ 
of $\mathfrak{g}$ according to 
\be
m_{\widehat{\lambda}}=\frac{|\lambda+\rho|^{2}}{2(k+g)}-\frac{|\rho|^{2}}{2g}.
\ee
For the singlet of $(\widehat{E}_{8})_{1}$ we have $k=1$, $\lambda=0$, $g=30$, $|\rho|^{2}=620$ and one 
recovers the quoted phase.
}.

From these relations one sees that it is possible to replace characters of $(\widehat{D}_{1})_{1}$ with characters of $(\widehat{D}_{5})_{1}$ while preserving modular invariance.
The precise mapping of characters  $(\widehat{D}_{1})_{1}\to (\widehat E_{8})_{1}\times (\widehat{D}_{5})_{1}$ 
is provided by the 
{\em Gepner map} 
\be
\label{eq:gepner-char}
(\id, \vector)\to \id\times (v, \id), \qquad (\spinor, \conjspinor)\to -\id\times (\5s, s).
\ee
Indeed, one can check that defining
\be
M = \left(\begin{array}{cccc}
0 & 1 & 0 & 0 \\
1 & 0 & 0 & 0 \\
0 & 0 & -1 & 0 \\
0 & 0 & 0 & -1
\end{array}\right),
\ee 
one has
\be
M\,S_{1}\,M = S_{5}, \qquad 
M\,T_{1}\,M = e^{-2\pi\,i/3}\,T_{5}.
\ee
The minus sign of fermionic characters has a double role. {On the one hand} it is required to fulfill modular $T$ invariance, {\em {\it i.e.}} level matching,
{and on the other hand} it enforces the spin-statistics condition {which requires} 
bosons and fermions to appear in the partition function with opposite multiplicities.
After the Gepner map states in a $(2,2)$ model have the structure $\Phi_{(2,2)}=\phi_{\C_\intern} \otimes \chi_{SO(10)}$. The construction is completed by two additional steps leading to well-defined spin-structures and space-time supersymmetry.

\underline{\em R/NS alignment}. Consistent quantization of the gauge fixed ${\cal N}=1$ supergravity theory requires
that the Ramond and NS sectors of the space-time and internal sectors are aligned.
After (\ref{eq:gepner-char}) this implies that $D_{5}$ spinor representations are aligned with the Ramond sector of the internal SCFT. Alignment  can be implemented by a SCMI that extends the chiral algebra by the current $J_{RNS}=J_v\otimes v$ (which has conformal weight $h_{RNS}=\frac{3}{2}+\frac{1}{2}=2$) because $Q_{J_v}\equiv 1/2$ for Ramond fields and 
$Q_{v}\equiv 1/2$ for $D_{5}$ spinors. Similarly, in the case of a Gepner model, where the internal SCFT  
${\cal C}_{int}=\bigotimes_{i} {\cal C}_{k_i}$ is a tensor product of $\N=2$ minimal models, the alignment can be implemented as a SCMI extending the chiral algebra by all bilinears of the respective supercurrents $J_{ij}=J_{v_i}J_{v_j}$, where $h_{ij}=3$.
In the following, we shall keep the alignment procedure explicit because we shall be interested in $(0,2)$ models for which the chiral algebra extension that implements the alignment only takes place in the right-moving sector, where it is needed for consistency.

\underline{\em Space-time supersymmetry}. We are interested in four dimensional space-time supersymmetry. Thus,  we want to perform a further projection
to a theory which admits a conserved 
supersymmetry charge exchanging bosonic and fermionic fields. 
In the string theory, this is nothing but a 
map between the Neveu-Schwarz and Ramond sectors. In $\N=2$ SCFT's, a natural
candidate is the total spectral flow operator, {\em {\it i.e.}} the simple current  $J_{GSO}=J_s\otimes s$. It has
integral conformal weight $h_{GSO}=c/24+5/8=1$ and hence
can be used for a SCMI of extension type.  Since {$Q_\GSO=-\frac{1}{2}q$, where $q$ refers to the $U(1)$ charge of a state $\Phi_{(2,2)}$,}
this {\em generalized GSO projection}
implies a projection to even $U(1)$ charges in the bosonic string and,
according to (\ref{eq:gepner-char}), to odd $U(1)$ charges in the Gepner construction
of the superstring \cite{Gepner} when the space-time contribution is taken 
into account.

As a final comment, we recall that the mechanism that implements space-time SUSY in the fermionic string is 
closely related, by the bosonic string map, to the mechanism that extends 
$E_8\times D_{5}$ to the gauge group $E_8\times E_6$ of a $(2,2)$ 
compactification. Indeed, the 33 massless vector bosons that extend the $45_{adj}$ of $D_5$ to
the $78_{adj}$ of the gauge group $E_6$ come from 
the $U(1)$ current of the $\N=2$ SCFT and  $2\times 16$ states associated with 
$(J_{GSO})^{\pm1}$.

\subsubsection{The $(0,2)$ model}

While $(2,2)$ models with 
$E_6$ gauge group can be constructed from a 4d bosonic string with internal CFT given by $\C_\intern \otimes SO(10) \times E_8$ after the Gepner map, the internal CFT needs to be split into smaller building blocks for $(0,2)$ models in order to be able to break supersymmetry only in the left-moving sector. We thus decompose $\C_\intern=\C' \otimes \F$, where $\C'$ is a general CFT while $\F$ is a minimal model at odd level $k=K-2$. In the Landau-Ginzburg phase  $\F$ has a Fermat-type superpotential $W=\Phi^{K}$ and hence will be referred to as \textit{Fermat factor}. In the gauge sector we start with an $SO(8)$ gauge group which we will then extend to $SO(10)$ in the left-moving bosonic sector and to $E_6$ in the right-moving sector which amounts to space-time supersymmetry after the Gepner map. Our $(0,2)$ models with $SO(10)$ gauge group hence are constructed from a 4d bosonic string with an internal $c=22$ CFT $\C'\otimes \F\otimes D_1\otimes D_4\times E_8$ with current algebras $D_n$ and $E_8$ at level 1 and a certain SCMI that will give rise to alignment of spin structures and the generalized GSO projection. States in a $(0,2)$ model then have the structure $\Phi_{(0,2)}=\phi_{\C'}\otimes \phi_{\F}\otimes \chi_{D_1}\otimes \chi_{D_4}$.\\
The SCMI that defines the resulting $(0,2)$ models is based on the simple current group generated by $J_{GSO},J_A, J_b,J_C$ with
\BE
	J_\GSO = J_s \otimes J_s \otimes s \otimes S,\qquad	J_A= 1 \otimes 1 \otimes v \otimes V,\qquad	J_C=  J_v \otimes 1 \otimes 1 \otimes V
\EE
and the Bonn twist 
\be
J_{b} = 1 \otimes (J_{s}^{K}\,J_{v}^{\frac{K-1}{2}})\otimes s\otimes 1
\ee
where the decomposition is with respect to $\C' \otimes \F \otimes D_1 \otimes D_4$ since $E_8$ acts as a spectator.\\
Charges in the Neveu-Schwarz  sector are quantized in units of $M'$ in $\C'$ and in units of $K$ in $\F$.   The central charge of the minimal model $\F$ is $c_{\F}=\frac{3k}{K}$ with $k=K-2$ while for the central charge of a general CFT $\C'$ we can only formally write $c' = \frac{3k'}{M'}$.~\footnote{While the value of the numerator $k'$, like the level $k$ of a minimal model, determines whether the universal center $\langle J_s,J_v\rangle$ with $J_s^{2M'}=J_v^{k'}$ and $J_v^2=\id$ is cyclic or not, the inverse charge quantum $M'$ is completely unrelated to this generalized ``level'' \cite{kr95,wkm}.}
Imposing $c'+c_{\F}=9$ we get $c'= 6\frac{K+1}K$. 
Since $k'=\frac{2M'(K+1)}{K}\in\IZ$ and $K$ is odd and relatively prime to $K+1$ 
we find that $K$ divides $M'$ and that $k'\in2\IZ$ and hence we get for the order of $J_\GSO$
\begin{equation} N_\GSO = \left\{\begin{array}{ll} 2M'	\qquad 
	\text{for} \quad M' \in 2\Z,\\
	4M'	\qquad \text{for} \quad M' \in 2\Z+1.\\
\end{array}\right.\end{equation} 
The orders of the alignment currents $J_A$ and $J_C$ are given by $N_A=N_C=2$ and the order of the Bonn twist is $N_b=4$. Notice, that $J_b^2=1 \otimes J_v \otimes v \otimes 1$ and hence the Bonn twist can be regarded as the square root of an alignment current. The order of our simple current group $\mathcal G$ is $N = 16M'$ for both, even and odd $M'$, because there is the relation $J_\GSO^{2M'} = J_b^2$ among the simple currents if $M' \in 2\Z+1$. Thus $\mathcal G$ can be parametrized by $J=J_{GSO}^\nu J_A^{\a}J_b^{\b}J_C^{\g}$ with $\a,\g=0,1$, $\b=0,1,2,3$ and $\nu=0,...,2M'-1$.\\
A SCMI as in (\ref{X}) is determined by the monodromy matrix $R$, as calculated from the simple current group $\mathcal G$, and the torsion matrix $X$, whose symmetric part $X_{(ij)}\equiv \frac{1}{2} R_{ij}$ is determined by $R_{ij}$ mod 1 for off-diagonal and by $R_{ii}$ mod 2 for diagonal elements while its antisymmetric part $X_{[ij]}\equiv X_{ij}-\frac{1}{2}R_{ij}$ corresponding to the discrete torsion in the orbifolding procedure is a priori subject to choice. Since the right-moving sector of our $(0,2)$ model is equivalent to that of a $(2,2)$ model we choose $X_{[ij]}$ such that we have full Ramond/Neveu-Schwarz alignment in the right-moving sector.
The non-vanishing monodromies between the simple currents $J_A$, $J_b$, $J_C$ and $J_\GSO$ are 
$R_{Ab} \equiv \frac{1}{2}$ mod 1 and $R_{bb} \equiv \frac{K-1}{2}$ mod 2. This fixes the symmetric part of $X$ and in addition we choose $X_{Ab}=\frac{1}{2}$ and $X_{bA}=0$. 
\begin{table}[!h]
\renewcommand{\arraystretch}{1.5}
\renewcommand{\tabcolsep}{0.2cm}
\begin{center}{\small
\begin{tabular}{||c||c|c|c|c\TVR4{1.5}||} \hline\hline
$R$     & $J_\GSO$& $J_A$ & $J_b$       & $J_C$ \\ \hline\hline
$J_\GSO$        & 0     & 0     & $0$ & 0 \\ \hline
$J_A$   & 0     & 0     & $\frac{1}{2}$  & 0 \\ \hline
$J_b$   & $0$ & $\frac{1}{2}$ & $\frac{K-1}{2}$    & 0 \\ \hline
$J_C$  & 0     & 0     & 0     & 0 \\ \hline\hline
\end{tabular}
\begin{tabular}{||c||c|c|c|c\TVR4{1.5}||} \hline\hline
$X$     & $J_\GSO$& $J_A$ & $J_b$       & $J_C$ \\ \hline\hline
$J_\GSO$        & 0     & 0     & 0  & 0 \\ \hline
$J_A$   & 0     & 0     &$\frac{1}{2}$ & 0 \\ \hline
$J_b$   & 0     & 0      & $\frac{K-1}{4}$ & 0 \\ \hline
$J_C$  & 0     & 0     & 0     & 0 \\ \hline \hline
\end{tabular}}
\caption{Monodromy matrix $R$ and torsion matrix $X$}
\label{table:montorsion}
\end{center}
\end{table}

\subsubsection{Generalized GSO projection and gauge/SUSY breaking for the $(0,2)$ model}
\label{sec:GSOandSUSY}

We want to construct heterotic string models with GUT gauge group $SO(10)$ and $\N=2$ supersymmetry only in the right-moving sector where it is needed to obtain space-time supersymmetry after the Gepner map to the heterotic string. This can be implemented by SCMI's that extend the left and right chiral algebra in an asymmetric way. 
Thinking of $[J] = \prod J_j^{\a_j}$ as the twist in the orbifolding procedure we can write the most general SCMI~\footnote{Notice that, as explained in \cite{ks93} 
and \cite{wkm}, one can also choose to work with $M_{i, [J]i}$ and impose projections on the left-moving states. Our choice, which projects right-moving states is motivated by the simpler structure of the right-moving sector where we have full R/NS alignment and better BPS properties.}
\BE
\label{SCMI}
M_{[J]i, i} = \mu(\Phi) \prod_j \delta_{Z} (\Qbar_j(\Phi) +  \alpha^k\, X_{kj}).
\EE
for a field $\PH_{[J]i, i}$ where the left-moving part is obtained by twisting the 
right-moving part with the  current $J=J_{GSO}^\nu J_A^\a J_b^\b J_C^\g$.
There are two types of invariants. Modular invariants of automorphism type are permutation matrices that uniquely map representation labels of the right-movers to the left-movers, where the permutation is an automorphism of the fusion rules. Let us define the kernel $Ker_{\Z} X$ as the set of integral solutions $[\a]$ of $ X_{kj}  \a^j \in \Z$
where $[\a]=[\nu,\a,\b,\g]$. If this kernel is trivial then $(\Qbar_j(\Phi) +  \alpha^k\, X_{kj})\in \Z$ has a unique solution $[\a]$ for each charge, which defines a unique position $[\a]\Phi$ on the orbit that only depends on the charge $\Qbar_i(\Phi)$. This yields an automorphism invariant. If $X=0$ we obtain a pure extension invariant because all fields with non-integral charges are projected out while all fields on an simple current orbit are combined to new conformal families. $X=0$ is only possible if the conformal weights of all simple currents $J \in \mathcal G$ are integral. 
Since these currents are in the orbit of the identity they extend the chiral algebras $\mathcal A_L$ and $\mathcal A_R$ so that we obtain a new rational symmetric and diagonal CFT.\\
Our $(0,2)$ model is given by $X\neq X^T \neq 0$ and is an asymmetric combination of an extension and automorphism type modular invariant partition function.
The extension of the right chiral algebra $\mathcal A_R$ is defined by  $Ker_{\Z} X$ which yields the extension $\tilde{\mathcal{A}_R} = \langle J_A, J^2_b, J_C, J_{GSO} \rangle$. We obtain the charge projection rules for the right-moving labels which amounts to the vanishing of all monodromy charges $\5Q_A\equiv \5Q_C \equiv\5Q_\GSO\equiv0$ modulo 1 except for $\5 Q_b \equiv \frac{\a}{2} + \frac{K-1}{4} \b$ modulo 1. From the form of $\tilde{\mathcal A_R}$ and from the charge selection rules we see that there is full alignment in the right-moving sector which justifies the choice of discrete torsion above.
Accordingly, the extension of the left-moving algebra is defined by $Ker_{\Z} X^T$, {\it i.e.} solutions $[\a]$ of $\a^k X_{kj} \in \Z$
, and yields $\tilde{\mathcal A_L}=\langle J_b, J_C, J_{GSO} \rangle$ for $K\equiv 5 \mod 4$ and $\tilde{\mathcal A_L}=\langle J_AJ_b, J_C, J_{GSO} \rangle$ for $K\equiv 3 \mod 4$.
As we will show below in more detail, the absence of the alignment current $J_A$ and the presence of the Bonn twist in the left chiral algebra already indicate that supersymmetry will be broken in the left-moving sector.

Since our asymmetric construction builds on a $D_4=SO(8)$ gauge group we need an extension mechanism to obtain a $D_5=SO(10)$ gauge group for the left-movers and a $E_6$ gauge group corresponding to space-time supersymmetry after the Gepner map for the right-movers.
Motivated by the free fermion construction of $D_n=SO(2n)$ in terms of $2n$ Majorana fermions with aligned spin structures where the extension of $SO(2m) \otimes SO(2n)$ to $SO(2m+2n)$ is implemented by a SCMI of extension type with the current $J=v_{D_m}\otimes v_{D_n}$, we will carry out an analogous "alignment extension" for our tensor product of (S)CFT's.

In the right-moving sector the extension is 2-fold. First we carry out an extension $D_1\otimes D_4\to D_5$ generated by the alignment current $J_A$ which is a prerequisite for a consistent Gepner map to the heterotic string. The further extension generated by $J_\GSO$ of $D_5 \rightarrow E_6$ on the bosonic version is then mapped to space-time SUSY on the heterotic side. 
On the left-moving side our class of models avoids the $J_A$ extension by an appropriate choice of discrete torsion (Table \ref{table:montorsion}) but uses the $J_{GSO}$ extension to promote the gauge group from $D_4$ to $D_5=SO(10)$.

\underline{\em Alignment extension.} 
The right-moving alignment extension $D_1\otimes D_4 \rightarrow D_5$ is generated by $J_A=v \otimes V$ with the charge projection $\Qbar_A=\Qbar_v+\Qbar_V$, where $\Qbar_v = \Qbar_V = 0$ for fields in the Neveu-Schwarz sector and $\Qbar_v = \Qbar_{V} = \frac{1}{2}$  for fields in the Ramond sector.\footnote{The group structure of character fusion is $\Z_{4}$ for odd $n$ and $\Z_{2}\times \Z_{2}$ for even $n$. The multiplication table is $sv=\5s$, $s^{2}=\5s^{2}=v^{n}$, $v^{2}=1$. }
Tensor products of fields from different sectors are projected out while tensor products of fields from the same sectors get combined to new conformal families with aligned spin structures. These are on the orbit generated by $J_A$ and read
\begin{equation}						
	\begin{array}{lllll}
\id& = &\id \otimes \id &\oplus& v \otimes V,\\
\vector& = &v \otimes \id &\oplus& \id \otimes V, \\
\spinor& = &s  \otimes S &\oplus& \sbar \otimes \Sbar,\\
\conjspinor& = &\sbar \otimes S &\oplus& s \otimes \Sbar,\\
\end{array}
\end{equation}
where $\vector, \spinor, \conjspinor$ denote characters in $D_5$ and the choice of $\spinor$ and $\conjspinor$ is convention. Notice that this $D_{1}\otimes D_{4}\to D_{5}$ extension is just a necessary step before applying the Gepner map and is not related to the $D_5=SO(10)$ gauge group in the left-moving sector which will be obtained by an extension using $J_\GSO$.
Due to our choice of discrete torsion the alignment current $J_A$ is not in the left chiral algebra and hence this alignment extension gets avoided in the left-moving sector.

\underline{\em GSO extension.} 
Contrary to the alignment current, $J_\GSO$ is in both, the left and right chiral algebra, yielding an extension of the gauge group to $SO(10)$ in the left-moving sector and a further extension of $D_5 \rightarrow E_6$ in the right-moving sector. The space-time part of the charge projection of $J_\GSO$, denoted by $Q_{\spinor}$, follows from the conformal dimensions $h_{\id} = 0, \ h_{\vector} = \frac{1}{2}, \ h_{\spinor} = h_{\conjspinor} = \frac{n}{8}$ and the monodromy formula (\ref{QJ}) yielding~\footnote{The formulas for the charge projection in the right-moving sector are given by the same expressions except for the exchange $Q_{\spinor} \leftrightarrow \Qbar_{\spinor}$.}
\be
  Q_{\spinor}(\id) = 0	\quad \text{and} \quad \ Q_{\spinor}(\vector) = \frac{1}{2}	
\ee
for fields in the NS sector and
\be
\label{eq:Qs}
 Q_{\spinor}(\spinor) = \frac{n}{4}-\left\{
\begin{array}{ll}
0 & n\in 2\Z \\
1/2 & n\in 2\Z+1
\end{array}
\right.
\quad \text{and} \quad
 Q_{\spinor}(\conjspinor) = \frac{n}{4}-\left\{
\begin{array}{ll}
1/2 & n\in 2\Z \\
0 & n\in 2\Z+1
\end{array}
\right.
\ee
for fields in the Ramond sector.
Due to the triality of the Dynkin diagram of $SO(8)$ the extension to $SO(10)$ based on $J_\GSO$ can be understood in terms of the alignment extension with a subsequent exchange of the characters $V \leftrightarrow S$ of $SO(8)$. 

\section{Computation of the charged massless spectrum}
\label{sec:CMS}

\subsection{Relation to the group theory and notation in DK and BW}

After restricting to the massless part of the spectrum, from the representation of the chiral algebras only the representation of the zero-mode algebras remain which, in the left-moving sector, contains the linearly realized subgroup $SO(10)\times U(1) \subset E_6$, where the abelian part is the absolutely defined $U(1)_{int}$ with charges $q_{int}$ of the internal $\N=2$ algebra $\C_{int}$. Likewise the quantum numbers of $SO(2)=D_1$ in the maximal subgroup $D_4\times D_1$ of $D_5=SO(10)$ are absolutely defined and we can identify the characters $\5s,\id,s,v$ of $SO(2)$ with the labels {$m=-1,0,1,\pm2$} in the conformal weights and $U(1)_m$ charges $(h_m,q_m)=(\frac{m^2}{8},\frac{m}{2})$ for primary fields $\Phi_m$, as introduced in \cite{bl96}. The label $m$ contributes to $Q_{GSO}$ with a prefactor $-\frac14$, as can be seen by evaluating, e.g., the contribution of $s=\Phi_{m=1}$ to the GSO charge projection, $Q_s^{D_1}(s)\equiv 2h_s -h_v \equiv -\frac14 \mod1$. With the $U(1)_m$ charges {$[-\frac12,0,\frac12,\pm1]$} of the characters $[\5s,\id,s,v]$, the $D_1$ charge $q^{D_1}$ contributes to $Q_{GSO}$ with a prefactor $-\frac12$ and we can write the charge projection by $J_\GSO$ as $0\equiv Q_\GSO \equiv-\frac12(q_{int}+q^{D_1})+Q^{D_4}_S \mod1$.

Translating this into the language of Distler--Kachru \cite{di94} and of Blumenhagen--Wi{\ss}kirchen \cite{bl95,bl96} can now identify the relevant $D_5=SO(10)$ decompositions under the maximal subgroup $SO(8)\times U(1)$
\begin{equation}
\begin{array}{llllllll}
\bf{1}& = & & &\bf{1}_0 & & & \\
\bf{10}& = & \bf{1}_{-2} &\oplus& \bf{8}^{s}_0 &\oplus& \bf{1}_2\\
\bf{16}& = & & \bf{8}^v_{-1} &\oplus& \bf{8}^{\5s}_{1}  \\
\bf{\overline{16}}& = & & \bf{8}^v_{1}  &\oplus& \bf{8}^{\5s}_{-1}. \\
\end{array}		\label{BW}
\end{equation}
The notation is $\mathbf{N}^{{\chi}}_{\widetilde q}$, where $N$ is the dimension of the $D_{4}$ representation, $\chi$ denotes the $SO(8)$ character and $\widetilde q=q_{int}+q^{D_1}$ is the $U(1)$ charge associated with the $U(1)$ current of the $SO(10)\supset SO(8) \times U(1)$ decomposition that is a linear combination of the $U(1)$ currents of the $\N=2$ algebra of $C_{int}$ and of $SO(2)=D_1$.

Given the values for $\widetilde q$ from the above decomposition and taking into account the space-time contribution coming from the $D_1$ factor we can determine the charge contribution $q_\intern$ from the internal sector. In the right-moving sector unitarity bounds highly restrict the values of $\5q_\intern$ and let us determine the spectrum of massless states completely. In fact, the right-moving states are all BPS states. The structure of $\N=2$ minimal models further enables us to derive bounds on the internal charge also in the left-moving  sector the states in which are on the orbit of the BPS states of the right-moving sector. This leaves us with a finite set of possible massless states of the heterotic $(0,2)$ string.

\subsection{Quantum numbers for chiral and vector multiplet}

After the alignment-extension of $D_4 \ex D_1$ to $D_5$ we can perform the
Gepner map on the right-moving side $D_5 \rightarrow D_1=SO(2)_\LC$ to obtain 
space-time quantum numbers (in light-cone gauge) from the representations of 
$D_5$.
The SUSY multiplets yielding space-time matter and space-time gauge symmetry generators are then assembled by $J_{GSO}$.
Admissible states are selected by imposing the massless condition\footnote{The NS vacuum in the right-moving sector has $\hbar=-\frac{1}{2}$.} $\hbar_{tot}=\hbar_\st+\hbar_\intern=\frac{1}{2}$ and the GSO projection $\overline Q_{GSO}\in \Z$ on the bosonized string. They are summarized in table \ref{chirvec} which in addition shows how the Gepner map $G$ acts on the characters of $D_{5}$ to get the associated space-time representation. From left to right we give the space-time conformal weight, the $D_{5}$ part of the GSO charge and the internal quantum numbers which are obtained by the charge selection rule $0 \equiv Q_\GSO \equiv -\frac{1}{2} \qbar_\intern + Q_s^{D_5}$ mod 1 and the unitarity bound $|\qbar_\intern|\le \frac{\5c}{6}= \frac{3}{2}$ for states in the Ramond sector and $|\qbar_\intern|\le 2\hbar_\intern$ for states in the NS sector.
\begin{table}[!h]
\begin{center}
{\renewcommand{\arraystretch}{1.5}
\renewcommand{\tabcolsep}{0.2cm}
\begin{tabular}{clc|c|c|c|c|c}
$\chi_{D_{5}}$ & $\stackrel{G}{\rightarrow}$ &$\chi_{SO(2)_{LC}}$ & $\overline{h}_{st}$ & $-2\,Q_{s}^{D_{5}}$ & $\overline{h}_{int}$ & $\overline{q}_{int}$ &  ${\rm state}$\\
\hline
$\id$   &$\rightarrow$ &$v$     & $\frac{1}{2}$ & $0$                 & $0$                  & $0$ & $\id$  \\
$\vector$    &$\rightarrow$& $\id$     & $0$                  & $1$                 & $\frac{1}{2}$ & $\pm 1$ & $\CP$, $\ACP$ \\
\hline
$\spinor$    &$\rightarrow$ &$- \5s$   &$\frac{1}{8}$  & $\frac{1}{2}$ & $\frac{3}{8}$ & $-\frac{1}{2}$ , $\frac{3}{2}$ & $\Rgs$ \\
$\conjspinor$ &$\rightarrow$ &$-s$   &$\frac{1}{8}$  & $-\frac{1}{2}$ & $\frac{3}{8}$ & $\frac{1}{2}$ , $-\frac{3}{2}$ & $\Rgs$
\end{tabular}}
\caption{Right-moving states with internal and space-time quantum numbers}
\label{chirvec}
\end{center}
\end{table}

Since on the right-moving side we have full R/NS alignment the $SO(2)_\LC$ representations are paired with internal states
of the same sector. From the condition for massless states and the unitarity bound it follows that the only admissible internal states are BPS states. In the NS sector the internal states that fulfill the BPS condition $\hbar_\intern=|\frac{\qbar_\intern}{2}|$
are chiral and antichiral states, denoted by $\CP$ and $\ACP$. In the Ramond sector the internal states that satisfy the analogous unitarity bound are Ramond ground states since $\hbar_\intern = \frac{\5c}{24}=\frac{3}{8}$ and are denoted by $\Rgs$. Note, that the  $\Rgs$ states with $\qbar_\intern=\pm \frac{3}{2}$ have  maximal/minimal $U(1)$ charge, respectively.

The SUSY multiplets are now assembled by $J_{GSO}$ as follows.
While the $D_5$ (or $SO(2)_{LC}$) representations are all on the same orbit, the
$U(1)$ charges $\qbar_{int}$ of the internal contribution to massless states are
shifted under spectral flow $J_s^{\C'}\otimes J_s^{\F}\subseteq J_{GSO}$ by 
$\pm \frac{3}{2}$, which quickly hits the unitarity bound $|\qbar_\intern|\le \frac{\5c}{ 6}$ for Ramond ground states and $|\qbar_\intern|\le 2\hbar_\intern$ in the NS sector.\\
Chiral multiplets consist of the lowest component of chiral superfields which are scalars with charge $\qbar_\intern=1$ (see table \ref{chirvec}) and their fermionic superpartners whose charge $\qbar_\intern=-\frac{1}{2}$ is shifted by spectral flow with respect to the scalars by $-\frac{3}{2}$ (a shift by $+\frac{3}{2}$ would yield a U(1) charge which is forbidden by the unitarity bound). Antichiral multiplets consist of the charge conjugate states of chiral multiplets.
Vector multiplets consist of the lowest component of vector superfields which are gauge bosons of charge  $\qbar_\intern=0$ and their superpartners which are left/right-handed gauginos of charge $\qbar_\intern=\pm \frac{3}{2}$. See \cite{Ka93} for a more detailed discussion.
The quantum numbers $(\bar h_{int},\bar q_{int})$ for the massless SUSY multiplets
hence are: 					
\vspace{-9pt}
\BI						\itemsep=0pt
\item	Vector multiplets: gauge bosons $(0,0)$ and left/right-handed gauginos
	$(3/8,\pm3/2)$.
\item	Chiral multiplets: fermions $(3/8,-1/2)$, scalars $(1/2,1)$
	and their charge conjugates.
\EI						\vspace{-9pt}

\subsection{Counting massless states}

In the right-moving sector the structure of massless states is highly constrained due to R/NS alignment following from supersymmetry while in the left-moving sector, where this alignment is partially broken, a broader range of possible states is admitted.
We can use the restricted structure in the right-moving sector and construct admissible left-moving states on orbits of admissible right-moving states, the pairings of which give the massless spectrum of the heterotic $(0,2)$ string.

In order to break supersymmetry only in the left-moving sector we have to start with smaller building blocks for the internal CFT as well as for the gauge group as discussed in section 2.3.2. Splitting $\C_\intern=\C'\otimes \F$ and starting with $SO(2) \times SO(8) \subseteq SO(10)$ we can write explicitly $h_\intern=h_{\C'}+h_\F$ and $h_{D_5}=h_{D_4}+h_{D_1}$.
Admissible left-moving states are obtained by twisting admissible right-moving states by $J=J_{GSO}^{\nu} J_A^{\a}J_b^{\b}J_C^{\g}$ with $\a,\g=0,1$, $\b=0,1,2,3$ and $\nu=0,...,2M'-1$
and imposing the condition for massless states $h_{tot} = h_{\C'} + h_\F + h_{D_1} + h_{D_4} = 1$ in the bosonic sector.~\footnote{The NS vacuum in the left-moving sector has $h=-1$.}
A generic left-moving state is obtained by a generic right-moving state by
\begin{equation}
|\C' \; \F \; D_1 \; D_4>_l = J_{GSO}^{\nu}\, J_A^{\a}\, J_b^{\b}\, J_C^{\g} \;|\C'\; \F\; D_1\; D_4>_r\\
\end{equation}
and the explicit form of the twist current is given by
\begin{equation}
\begin{array}{lllllllll}
J &=& J_s^\nu J_v^{\g} &\otimes& J_s^{\nu+\b K} J_v^{\frac{K-1}{2}\b} &\otimes& 
s^{\nu+\b} v^{\a} &\otimes& S^\nu V^{\a+\g}.
\end{array}
\end{equation}
Besides organizing the contributions to the spectrum in twisted sectors the exponents $\nu,\a,\b,\g$ determine whether a left-moving (twisted) factor yields the same field as the right-moving factor on the orbit of which it is computed or its superpartner. By choosing a specific SUSY multiplet together with an $SO(10)$ representation for the gauge multiplet we will study the structure of the charged massless spectrum of non-singlet matter states. 
We will use the information obtained from the exact CFT calculations to determine the number of generations, antigenerations and vectors by the means of the extended Poincar\'{e} polynomial and the {complementary Poincar\'{e}} polynomial.

\subsubsection{The extended Poincar\'{e} polynomial}

The EPP of a $N=2$ SCFT as given by\cite{kr95}
\be
P(t, \ovl t, x) = \sum_{l\ge 0}\sum_{\k=0,1} x^l\,(-1)^\k\, P_{l,\k}(t,\ovl t),
\ee
 is the sum of $l$  $x$-twisted Poincar\'{e} polynomials weighted by an additional change of sign, that is related to the ambiguity of dealing with a field or its superpartner. The ordinary Poincar\'{e} polynomial is given by~\footnote{Note, that the EPP is conventionally defined with right-movers $\5a$ on the orbit of left-movers $a$ whereas in our analysis we choose left-movers on the orbit of right-movers due to the nicer BPS structure in the right-moving sector.}
\be
P_{l,\k}(t,\ovl t) = \mathop{\mathop{\sum_{( a,  \5a) \in {\cal R}_{(c,c)}}}_{
\5a = J_s^{2\,l}\,J_v^\k\, a}} t^{q(a)}\,{\ovl t}^{\5q(\5{a})},
\ee
where the sum is over states in the $(c,c)$ ring.
In the case where the internal sector has aligned spin structures (corresponding to a twist by an even exponent of $J_b$) the states contributing to the massless spectrum are BPS states. We can determine the number of generations, antigenerations and vectors by looking for particular terms in the EPP that are determined by the $U(1)$ charges of the internal left- and right-moving sector as will be calculated below.

\subsubsection{The {complementary Poincar\'{e}} polynomial}

In the case where the internal sector has non-aligned spin structures (twist by an odd exponent of $J_b$) also non-BPS states can contribute to the massless spectrum and we thus need in addition to the information of the (left-moving) internal $U(1)$ charge also the conformal weight.
We are thus interested in the {complementary Poincar\'{e}} polynomial
\be
\label{eq:P}
\mathscr{P}(x, q, t) = \sum_{\ell\ge 0}\sum_{k=0,1}\mathop{\sum_{\5a\in{\cal R}_{0}}}_{a =
 J_{s}^{2\ell} J_{v}^{k} \5a}
(-1)^{k}  \, x^{\ell}\, q^{H_{L}(a)} t^{Q(a)},
\ee
where $\5a$ runs over the Ramond ground states and the sum over $a$ is over 
\underline{all states} (including descendants)
in the conformal family of $J_{s}^{2\ell} J_{v}^{k} \5a$~\footnote{The limit $q=0$ equals the EPP at  $\5t=1$
and with $t\to 1/t$. This replacement is necessary since we defined the CPP in (\ref{eq:P}) with left-movers $a$ on the orbit of right-movers $\5a$.}.
This polynomial is {complementary} to the EPP. It does not involve the right-mover's charge, but instead 
keeps track of the conformal dimension of excited left-moving states.

We can compute $\mathscr{P}(x, q, t)$ in terms of the elliptic genus which, 
for a general $\N=2$ SCFT, is the trace \cite{w93}�
\be
\label{eq:elliptic-genus}
Z(q, \5q, t) = {\rm Tr}_{\cal H} (-1)^{F} q^{H_{L}} \5q^{H_{R}}\,t^{Q_{L}},
\ee
where ${\cal H}$ is the full Hilbert space, $H_{L,R}$ are the Hamiltonians of left- and right-movers, 
$Q_{L}$ is the $U(1)$ charge of left-movers, and 
$F = F_{L}+F_{R}$ is the total fermion number~\footnote{As usual, we can identify $(-1)^{F_{L,R}}
= e^{i\pi\,Q_{L,R}}$.}. Up to spectral flow, we can assume that the left- and right-movers are
in the  Ramond sector.
By supersymmetry, the non vanishing contributions to $Z$ come from the 
states where the right-mover is a  ground state $H_{R}=0$ and thus $Z(q,\5q, t) = Z(q, 1, t) \equiv Z(q, t)$.
As discussed in  \cite{w93}, the elliptic genus of a Landau-Ginzburg model can be computed in free field theory. 
{Let the superpotential $W(\Phi_{1}, \dots, \Phi_{N})$ be a holomorphic function in the chiral superfields $\{\Phi_i\}_{i=1, \dots, N}$ 
such that 
\be
W(\lambda^{\omega_{1}} \Phi_{1}, \dots, \lambda^{\omega_{N}} \Phi_{N}) = \lambda\,W(\Phi_1, \dots, \Phi_{N}).
\ee}
Then, we can write $Z_{\rm LG}(q, t) = \prod_{i} Z_{\omega_{i}}(q, t)$ with~\footnote{{Our notation is related to \cite{w93} by $t=e^{i\g}$ and $w=\a$ and to that of \cite{kaw93a} by $t=y=e^{2\pi i z}$.}}
\be
\label{eq:single}
Z_{\omega}(q, t) = t^{-\frac{1-2\omega}{2}}\,\frac{1-t^{1-\omega}}{1-t^{\omega}}
\prod_{n=1}^{\infty} \frac{1-q^{n}\,t^{1-\omega}}{1-q^{n}\,t^{\omega}}
\frac{1-q^{n}\,t^{-(1-\omega)}}{1-q^{n}\,t^{-\omega}} = 
 \frac{\vartheta_{1}(q, t^{1-\omega})}{\vartheta_{1}(q, t^{\omega})},
\ee
where the Jacobi theta function $\vartheta_{1}$ is given by
\be
\vartheta_{1}(q, t) = i\,\sum_{n\in\Z}(-1)^{n} q^{\frac{1}{2}(n-\frac{1}{2})^{2}}\,t^{n-\frac{1}{2}}.
\ee
The expression (\ref{eq:single}) is obtained immediately in free field theory. One just keeps track of the contributions 
of the scalar $\phi$ and left-moving fermion $\psi_{-}$ in $\Phi$ as well as their complex conjugates \cite{w93}.
The polynomial (\ref{eq:P}) is simply the sum over the twists along the spatial direction. Notice that the change of 
 sign $(-1)^{k}$ in (\ref{eq:P}) due to $J_{v}$ applications is automatically
taken into account by the fermion sign $(-1)^{F}$ in the elliptic genus. The effect of the spatial twist can be obtained
by standard orbifold techniques and gives the contribution \cite{day94}
\be
q^{\frac{\widehat c}{2}\ell^{2}}\,t^{\widehat{c} \ell}\,
Z_{\rm LG}(q, q^{\ell}\,t),
\ee
with $\widehat{c} = \frac{c}{3} = \sum_{i=1}^{N}(1-2\omega_{i})$.
The sum over $\ell$-twisted factors can include phases as usual in orbifold partition functions \cite{day94}. The choice of 
trivial phases reproduces the EPP at $\5t=1$ in the $q\to 0$ limit and we simply obtain
\be
\mathscr{P}(x, q, t) =  \sum_{\ell\ge 0} x^{\ell}\,q^{\frac{\widehat c}{2}\ell^{2}}\,t^{\widehat{c} \ell}\,
Z_{\rm LG}(q, q^{\ell}\,t).
\ee
As a check, a tedious exercise (see appendix)  gives indeed
\be
\lim_{q\to 0} \mathscr{P}(x, q, t) = P(t^{-1}, 1, x)
\ee

\subsection{Counting generations}
\label{sec:countingGEN}

We have now assembled all tools that we need in order to compute the charged massless spectrum.
As a representative of the right-moving sector we consider a space-time matter scalar.
From table \ref{chirvec} we can read off the right-moving internal conformal weight and charge to be $\5h_\intern=\frac 12$ and $\5q_\intern=\pm1$ which corresponds to a chiral/antichiral state in the internal sector. Let us consider the antichiral state. Splitting $\C_\intern=\C' \otimes \F$ we can write the right-moving state as~\footnote{This choice of right-moving representative forbids to further use the field identifications (\ref{fieldID}) for left-movers, since those must be applied simultaneously on both sides in order to yield an admissible state that contributes to the spectrum of the heterotic string. Field identifications that are based on the modular properties of the labels, however, can still be used.}
\be
\label{STscalar}
\Psi_{\rm right} = |\C'\otimes \F\,; \,D_{1}\otimes D_{4}\rangle_{r} = |\5\Phi \otimes \5\varphi^{\5\ell, 0}_{\5\ell}\,;\, \id\otimes V\rangle_{r}
\ee
with $\5\ell = 0, \dots, K-2$. With $K$ being the charge quantum in the NS sector we can explicitly compute the charge of the antichiral state $\5\varphi^{\5\ell, 0}_{\5\ell}$ in the Fermat sector and, hence, we can split $\5q_\intern$ into contributions form $\F$ and $\C'$ according to 
\be
\5q^{F}_{\ACP} = -\frac{\5\ell}{K}  \qquad \text{and} \qquad \5q'_{\ACP} = \frac{\overline{\ell}-K}{K}.
\ee
The charge projection rules $\5Q_A\equiv \5Q_C\equiv \5Q_\GSO \equiv 0\mod 1$ for the right-movers are already satisfied by (\ref{STscalar}). The monodromy charge $\5Q_b(\Psi_{\rm right}) \equiv h_b+h_{\Psi_{\rm right}}-h_{J_b\Psi_{\rm right}}$ involving the Bonn twist can be computed to yield
\be
\5Q_b \Psi_{\rm right} \equiv \left\{\begin{array}{ll}
0       & \quad \overline{\ell}\in 2\Z      \\
1/2         & \quad \overline{\ell}\not\in 2\Z .
\end{array}\right.
\ee
Comparing this result to the projection rule $\5Q_b\equiv \frac{\a}{2}+\frac{K-1}{4} \b \mod1$ obtained in section \ref{sec:GSOandSUSY}, restricts the possible exponents $\a,\b,\g,\nu$ of $J=J_A^{\a} J_b^{\b} J_C^{\g} J_\GSO^{\nu}$ by which the admissible right-moving states are twisted to yield admissible left-moving states.\\
Since we want to count generations, as represented by states in ${\bf 16}$, the left-moving states must transform under $SO(8)\times U(1)$ as $\mathbf{8}^{v}_{-1}$ or $\mathbf{8}^{\5s}_{1}$ according to (\ref{BW}). For convenience we stay in the NS sector where states are of the general form 
\be
\Psi_{\rm left} = |\C' \otimes \F\,;\, D_{1}\otimes V\rangle_{l}.
\ee
With the massless condition $h_{int}+h^{D_{1}}+\half = 1$ and the charge condition $q_{int}+q^{D_{1}} = -1$, as follows from the group theory discussion, we have four possibilities for admissible left-moving states. Their space-time parts, conformal weights and $U(1)$ charges are   
\be
{\renewcommand{\arraystretch}{2.2}
\renewcommand{\tabcolsep}{0.08cm}
\begin{array}{lll}
\label{allleftgen}
&|\id \otimes V\rangle_{l} \quad \text{with} \quad h_\intern=\frac12\, ,\;q_\intern=-1; \qquad
&|v \otimes V\rangle_{l} \quad \text{with} \quad h_\intern=0 \, ,\; q_\intern=-\frac32\\
&|s \otimes V\rangle_{l} \quad \text{with} \quad h_\intern=\frac38 \, ,\; q_\intern=-\frac32; \qquad
&|\5s \otimes V\rangle_{l} \quad \text{with} \quad h_\intern=\frac38 \, ,\; q_\intern=-\frac12.
\end{array}}
\ee
While in the right-moving sector all factors are aligned, in the left-moving sector this alignment is partially broken due to the presence of the Bonn twist $J_b$ in the extension of the left-chiral algebra. The remaining alignment between the factors $\C'$ and $D_4$ is due to the current $J_C$.
Depending on whether the two factors $D_1$ and $D_4$ and, hence, also $\C'$ and $\F$ are aligned or not, there is a qualitatively different analysis for counting the number of generations.

\subsubsection{Aligned Generations}

For an even power of the Bonn twist the internal factors $\F$ and $\C'$ and, hence, also $D_1$ and $D_4$ are aligned along the orbit generated by the twist. This case corresponds to the states in the first line of (\ref{allleftgen}). However, taking into account the BPS bound $h_\intern \geq \frac{|q_\intern|}{2}$, states with $q_\intern=-\frac32$ cannot appear in the massless spectrum. States with $q_\intern=-1$ do satisfy the bound and, furthermore, the internal part is an antichiral primary state. Hence, the only admissible left-moving states with aligned factors are of the form~\footnote{Note, that admissible left-moving states in the $\F$ sector could in principle also appear as $\varphi^{K-2-\ell,2}_{\ell+K}$ which is dual to $\varphi^{\ell,0}_{\ell}$ under field identification. From the previous footnote, however, it follows that after having fixed the right-moving representative we cannot use field identifications (\ref{fieldID}) on the left-moving side anymore, and hence we have to discuss both possibilities. Since their $U(1)$ charges are equal we can cover both cases in one shot by taking into account the two possible labels $\ell$ and $K-2-\ell$ when counting generations via the EPP.}
\be
\label{Leftaligned}
\Psi_{\rm left} = |\Phi \otimes \varphi^{\ell,0}_{\ell}\,;\, \id \otimes V\ket_l \qquad \text{with} \quad h_\intern=\frac12\, ,\;q_\intern=-1
\ee
with $\ell=0,\dots,K-2$. The $U(1)$ charge of the antichiral primary state $\varphi^{\ell,0}_{\ell}$ in the Fermat factor can be computed and, hence, the charge contributions from the $\F$ and $\C'$ sectors to $q_\intern=-1$ are
\be
q^{F}_{\ACP} = -\frac{\ell}{K} \qquad \text{and} \qquad q'_{\ACP} = \frac{\ell-K}{K}.
\ee
The left-moving state (\ref{Leftaligned}) is on the orbit of the right-moving state (\ref{STscalar}) if
\be
\Psi_{\rm left}=J_A^{\a} J_b^{\b} J_C^{\g} J_\GSO^{\nu} \Psi_{\rm right}.
\ee
Explicitely, this means
\ba
\label{eq:orb1}
\Phi &=& J_{s}^{\nu}J_{v}^{\gamma}\, \5\Phi, \\
\label{eq:orb2}
\varphi^{\ell, 0}_{\ell} &=& J_{s}^{\nu+K\beta} J_{v}^{\beta\frac{K-1}{2}}\,\5\varphi^{\5\ell, 0}_{\5\ell}, \\
\id &=& s^{\nu+\beta} v^{\alpha}, \\
V &=& S^{\nu} V^{\alpha+\gamma+1}.
\ea
Using the fusion rules~\footnote{Indeed, the fusion rules imply that any monomial in $s$, $v$, and $\5s$ can be reduced to $s^{p}$ which is $\id$ iff $p \equiv 0\mod 4$, in agreement with the $\Z_{4}$ structure. In the case of $D_{4}$, any monomial can be reduced to the form $S^{p}V^{q}$ which is $\id$ iff $p$ and $q$ are even, in agreement with the $\Z_{2}\times \Z_{2}$ structure.}, the last two equations read
\ba
\label{constraints}
\nu &\in& 2\Z, \nonumber \\
\alpha+\gamma &\in& 2\Z, \\
\nu+\beta+2\alpha &\equiv& 0 \mod 4.\nonumber
\ea
These constraints together with the charge projection rule
\be
\label{QB}
\5 Q_b \equiv \frac{\a}{2} + \frac{K-1}{4} \b \equiv \left\{\begin{array}{ll}
0       & \quad \overline{\ell}\in 2\Z      \\
1/2         & \quad \overline{\ell}\not\in 2\Z .
\end{array}\right. \mod 1
\ee
uniquely determine the possible combinations of twist exponents $\a,\b,\g,\nu$ such that the combination of $\Psi_{\rm right}$ and $\Psi_{\rm left}=J_A^{\a} J_b^{\b} J_C^{\g} J_\GSO^{\nu} \Psi_{\rm right}$ give contributions to the massless spectrum.\\
In order to count generations we need to find the appropriate terms in the extended Poincar\'{e} polynomial. We use the simple structure in the Fermat sector in order to determine the admissible terms for the EPP in the $\C'$ sector. For a minimal model, like $\F$, the EPP over the chiral ring is given by \cite{kr95}
\be
P_{(c,c)}(t^{K}, \ovl t^{K}, x) = \sum_{l=1}^{K-1} (t\,\ovl t)^{l-1}\,\frac{1-(-x)^l\,{\ovl t}^{K-2\,l}}{1-(-x)^{K}}.
\ee
In general, the order of the spectral flow in $\C'$ can be larger than that in $\F$. Therefore, we need the full 
'periodic' expansion of the EPP with arbitrarily high powers of $x$, which, because of $K\not\in 2\Z$, reads 
\be
\label{fullperiodicEPP}
P_{(c,c)}(t^{K}, \ovl t^{K}, x) = \sum_{l=1}^{K-1} (t\,\ovl t)^{l-1}\,\Big(1-(-x)^l\,{\ovl t}^{K-2\,l}\Big) \sum_{r=0}^{\infty}(-1)^{r} x^{rK}.
\ee
There are two distinct types of terms in $P_{(c,c)}(t, \ovl t, x)$, which, after identification of the exponents of $t$ and $\5t$ with the charges $q_F=-\frac{\ell}{K}$ and $\5q_F=-\frac{\5\ell}{K}$ of states in the $\F$ sector, read~\footnote{Notice, that since the states in the Fermat sector are antichiral states we actually have to sum over the antichiral ring in (\ref{fullperiodicEPP}) which simply amounts to adding a factor $(t\5t)^{-\frac{Kc_\F}{3}}=(t\5t)^{2-K}$.}
\be  \F: \quad \begin{array}{lll}
(i) & (-1)^{r} \,\ \ \ \ x^{rK}\ \ \  \ \ \ \,t^{-\frac{\ell}{K}} \, \5t^{-\frac{\5\ell}{K}} \qquad &\ell=\5\ell,\\
(ii) & (-1)^{r+\5\ell}\,\ x^{\5\ell+1+rK}\, \ t^{-\frac{\ell}{K}}\,\5t^{-\frac{\5\ell}{K}}  \qquad &\ell=K-2-\5\ell.
\end{array} \ee
For each admissible combination of $\a,\b,\g$ and $\nu$ (as follows from (\ref{constraints})) the label $\5\ell$ and the parameter $r$ can be determined by the charge projection rule (\ref{QB}) and by comparison of (\ref{eq:orb2}) with the generic structure of terms in the EPP of $\F$ above.\footnote{If there is no supercurrent in (\ref{eq:orb2}), admissible terms have positive coefficient otherwise they have a negative coefficient. For $n=\frac{\nu}{2} \in 2\Z$ the exponent of $x$ must be even and otherwise odd.} 
For each of these combinations we can then determine the generic structure of admissible terms in the EPP over the chiral ring of $\C'$ yielding
\be  \C': \quad \begin{array}{ll}
(i) & (-1)^{\g} \,\ \ \ x^{rK}\ \ \  \ \ \ \,t^{q'_{\CP}} \, \5t^{\5q'_{\CP}}\\
(ii) & (-1)^{\g}\,\ \ \ x^{\5\ell+1+rK}\, t^{q'_{\CP}}\,\5t^{\5q'_{\CP}},
\end{array} \ee
with $\5\ell$ and $r$ being determined by the admissible terms in the EPP of $\F$. 
Note, that the sign $(-1)^{\g}$ depends on the exponent of $J_C$ because it determines whether or not a supercurrent is applied to states in $\C'$, as follows from (\ref{eq:orb1}). The admissible terms in the EPP $P_{(c,c)}(t, \ovl t, x)$ of $\C'$ are summarized in table \ref{Agen}. 
Note, that upon spectral flow the $U(1)$ charge of the internal antichiral primary states is shifted to that of chiral primary states by $\frac{c'}{3}=\frac{2K+2}{K}$ and we parametrized the exponent of $J_\GSO$ by $\nu=2n$. In table \ref{Agen} the data necessary for counting aligned generations is collected. 
\begin{table}[!h]
\begin{center}
{\renewcommand{\arraystretch}{1.5}
\renewcommand{\tabcolsep}{0.3cm}
{\small \begin{tabular}{||c|c|c|c|c||} \hline \hline
\multicolumn5{||c||}{\bf 16 - aligned generations}\\ \hline \hline
$\sigma'$ 	& $\5\ell$	& l	 			& $\5q'_{\CP}$ 			& $q'_{\CP}$ \\ \hline \hline
$+$		& $\5\ell \in 2\Z, \quad  0\leq \5\ell \leq k$		& $\begin{array}{ll} l\in 2K\Z \quad &\b=0, \; n \in 2\Z\\  l\in2K\Z+1 \quad &\b=2, \; n \notin 2\Z \end{array}$	& $\frac{K+2+\5\ell}{K}$	& $\frac{K+2+\5\ell}{K}$\\ \hline
$-$		& $\5\ell \notin 2\Z,  \quad 0\leq \5\ell \leq k$	& $\begin{array}{ll} l\in 2K\Z+1 \quad &\b=0, \; n \notin 2\Z\\  l\in 2K\Z \quad &\b=2, \; n \in 2\Z \end{array}$	& $\frac{K+2+\5\ell}{K}$	& $\frac{2K-\5\ell}{K}$\\
\hline \hline
 \end{tabular}}}
\caption{Left- and right-moving $\C'$-sector charges $q'_{\CP}$ and $\5q'_{\CP}$, right-moving label $\5\ell$, exponent $l$ of $x$ and sign $\sigma'=(-1)^{\g}$ in the EPP of $\C'$ with terms $~\sim \sigma' \ x^l\ t^{q'}\ \5t^{\5q'}$.}
\label{Agen}
\end{center}
\end{table}

\subsubsection{Non-aligned Generations}

For an odd power of the Bonn twist the Fermat sector and the $\C'$ sector are not aligned anymore. This case corresponds to the states in the second line of (\ref{allleftgen}). Due to the remaining alignment by $J_C$ states in $\C'$ are in the NS sector while states in $\F$ are in the Ramond sector. 
By exploiting the well known structure  of the Fermat sector we can gain some insight in which states in $\F$ and, hence, in $\C'$ are admissible and which are not.
The fields in $\F$ are $\Ph_m^{\ell s}\sim\Ph_{m+K}^{k-\ell,s+2}$ with $0\le \ell\le k$,
$m$ mod $2K$, $s$ mod 4 and $\ell+m+s\equiv0$ mod 2, {\it i.e.} there are
$2K(K-1)$ fields. The `generic' subgroup of the
center $\langle J_s=\Ph_1^{01},J_v=\Ph_0^{02}\rangle$ has order $4K$.
Because of field identification its orbits are labeled by $\ell\sim k-\ell$ and
there are fixed points (with $\ell=k/2$) precisely for even level $k$.
Each orbit of the generic center contains two BPS states of each type
($\CP$, $\ACP$ and $\Rgs$).
The left-moving states $\phi^{ls}_m$ on the orbit of $\ovl \phi^{\5\ell,0}_{\5\ell}$ are at position $J_s^{m-\ell}J_v^{\frac{s-m+\ell}{2}}$ with $\ell=\5\ell$~\footnote{Notice, that the label $\ell$ of $\varphi^{\ell,s}_m$ does not change under fusion with $J_s$ or $J_v$. See section \ref{sec:MM}.} and $s=\pm1$ (Ramond sector).
Since $h'+h_{\F} = \frac38$ the Ramond sector states in $\F$ must have $h_{\F} \leq \frac38$. A straightforward analysis shows that the BPS bound $h'\geq\frac{|q'|}{2}$ is not satisfied by states with $q_\intern=-\frac32$. 
Hence, the only admissible left-moving states with non-aligned $\F$ and $\C'$ factors are of the form
\be
\label{leftstateNA}
\Psi_{\rm left} = |\Phi \otimes \varphi\,;\, \5s \otimes V\ket_l \qquad \text{with} \quad h_\intern=\frac38\, ,\;q_\intern=-\frac12.
\ee
It is  on the orbit of $\Psi_{\rm right}$ if
\be
\Psi_{\rm left} = J_{GSO}^{\nu} J_{A}^{\alpha} J_{B}^{\beta} J_{C}^{\gamma} \Psi_{\rm right},
\ee
which explicitely reads
\ba
\label{eq:orb1NA}
\Phi &=& J_{s}^{\nu}J_{v}^{\gamma}\, \5\Phi, \\
\label{eq:orb2NA}
\varphi^{\ell,s}_m&=& J_{s}^{\nu+K\beta} J_{v}^{\beta\frac{K-1}{2}}\,\5\varphi^{\5\ell, 0}_{\5\ell}=\varphi^{\5\ell,\nu+2\b K-\b}_{\5\ell+\nu+\b K}, \\
\5s&=& s^{\nu+\beta} v^{\alpha}, \\
V &=& S^{\nu} V^{\alpha+\gamma+1}.
\ea
Using the fusion rules the last two equations read
\ba
\label{eq:discrete-odd}
\nu &\in& 2\Z, \nonumber \\
\alpha+\gamma &\in& 2\Z, \\
\nu+\beta+2\alpha &\equiv& 3 \mod 4 \nonumber.
\ea
Together with the charge projection from above
\be
\5 Q_b \equiv \frac{\a}{2} + \frac{K-1}{4} \b \mod 1 \equiv \left\{\begin{array}{ll}
0       & \quad \overline{\ell}\in 2\Z      \\
1/2         & \quad \overline{\ell}\not\in 2\Z .
\end{array}\right. \mod 1 
\ee
the constraints (\ref{eq:discrete-odd}) strongly restrict the possible combinations of twist exponents and labels as would follow from (\ref{eq:orb2NA}). Using $\ell=\5\ell$ and the BPS condition we can now determine which left-moving states in the Fermat sector lead to massless states.
In the simplest case they are Ramond ground states. Those of the form $\varphi^{\5\ell, 1}_{(\5\ell+1)}$ need to satisfy the lower bound $\5\ell \geq \frac{K-3}{2}$, those of the form $\varphi^{\5\ell,- 1}_{-(\5\ell+1)}$ must satisfy the upper bound $\5\ell \leq \frac{K-1}{2}$.  Their superpartners $\varphi^{\5\ell,\mp 1}_{\pm(\5\ell+1)}$ are not admissible since their conformal weights are $h=\frac{c_\F}{24}+1>\frac38$. Apart from ground states there are excited Ramond states of the form $\varphi^{\5\ell,-1}_m$ with $s=-1$ and $|m|<\5\ell$ in the standard range where they have to satisfy the condition $(|m|-1)^2\geq \5\ell(\5\ell+2)+1-K$, as follows from the BPS bound, the branching rule $\5\ell+m\equiv 1\mod2$ and the simple upper bound $\5\ell\leq \frac{K+3}{6}$. Furthermore, there are excited Ramond states of the form $\varphi^{\5\ell,1}_m$ with $|m|>\5\ell$ outside the standard range. We can always bring these states back to the standard range via field identifications where they need to satisfy $s+2=\pm1$, the branching rule $\5\ell+m \equiv 1\mod2$ and the conditions $|m\pm K|\leq K-2-\5\ell$ and $(|m\pm K|-1)^2 \geq (K-\5\ell-2)(K-\5\ell)+1-K$.
After having determined all admissible states in the Fermat sector with conformal weights and $U(1)$ charges as calculated from (\ref{hqMM}) we can compute the conformal weights and charges of admissible states in the $\C'$ sector.
The novelty, as compared to the case of aligned generations, is that also non-BPS states in $\C'$ can contribute to the spectrum of massless states. Therefore, in addition to the left-moving $U(1)$ charge in $\C'$ we also need to keep the information of the left-moving conformal weight.~\footnote{The right-moving charge is not necessary. Suppose that our tables find a candidate left-moving state in $\C'$ along the orbit of the various currents. This means that the right-moving $\C'$ state is a Ramond ground state (after spectral flow) with charge $\5q'$ obeying $\5q'  \mod 2 = \frac{\5\ell+1}{K}$ with $\5\ell = 0, 1, ..., K-2$. However, $\5q'$ must be such that $|\5q' | \le c'/6 = 1+1/K$. This easily shows that $\5q'=\frac{\5\ell+1}{K}$ exactly, {\it i.e.} without mod 2.}
In order to count generations we are, hence, looking for admissible terms in the {complementary Poincar\'{e}} polynomial in the $C'$ sector which is, in some sense, {complementary} to the extended Poincar\'{e} polynomial. As follows from (\ref{eq:P}), admissible terms in the {complementary Poincar\'{e}} polynomial have the generic structure
\be
(-1)^{\g} \ x^{\nu/2}\ q^{h_R'-\frac{c'}{24}} \ t^{q_R'},
\ee
with the $U(1)$ charges and conformal weights in the Ramond sector. The sign $\sigma'=(-1)^{\g}$ is determined by (\ref{eq:orb1NA}), {\it i.e.} whether or not a supercurrent is applied, and the exponent of $x$ can be read off by comparison of (\ref{eq:P}) with (\ref{eq:orb1NA}) to be $\nu/2$.~\footnote{While the information on the exponent of the GSO current can be determined modulo $4K$, all $\nu$ within the range of $1,...,N_\GSO$ need to be considered separately in the counting algorithm. In other words, values of $\nu$ that are absolutely different but the same modulo $4K$ generically appear as exponent of different terms with different coefficients in the {complementary Poincar\'{e}} polynomial.} The admissible terms in the {complementary Poincar\'{e}} polynomial of $\C'$ are summarized in table \ref{NAgen} which is organized in terms of the left-moving Fermat sector states $\varphi^{\ell,s}_{m}$ as depicted in the first column.
\begin{sidewaystable}
\begin{center}
{\renewcommand{\arraystretch}{2.2}
\renewcommand{\tabcolsep}{0.08cm}
{\scriptsize \begin{tabular}{||c|c|c|c|c|c|c|c||} \hline \hline
 \multicolumn8{||c||}{\bf 16 - non-aligned generations}\\ \hline \hline
Fermat						& $K \mod 4$ 	& $\sigma'$ 	& $\5\ell$				
&$m$			& $\nu \mod 4K$			& $h_R'$ 			& $q_R'$\\ \hline \hline

 $\varphi^{\5\ell,1}_{\5\ell+1}$		& $1$		& $-$		& $\5\ell \notin 2\Z, \quad \frac{K-3}{2} \leq \5\ell \leq k$	
& $m=\5\ell+1$ 		& $1-\b K$			& $\frac{\5\ell+2}{2K}$		& $\frac{\5\ell+2}{K}$\\ \hline
						& $3$		& $-$		& $\5\ell\in 2\Z, \quad \frac{K-3}{2} \leq \5\ell \leq k$
& $m=\5\ell+1$ 		& $1-\b K+2K$		 	& $\frac{\5\ell+2}{2K}$		& $\frac{\5\ell+2}{K}$ \\\hline

 $\varphi^{\5\ell, -1}_{-(\5\ell+1)}$		& $1$		& $+$		& $\5\ell\in 2\Z, \quad 0 \leq \5\ell \leq \frac{K-1}{2}$
& $m=-(\5\ell+1)$	& $-1-2\5\ell-\b K$		& $\frac{K-\5\ell}{2K}$		& $\frac{K-\5\ell}{K}$\\ \hline
						& $3$		& $+$		& $\5\ell \notin 2\Z, \quad 0 \leq \5\ell \leq \frac{K-1}{2}$ 	
& $m=-(\5\ell+1)$	& $-1-2\5\ell-\b K$		& $\frac{K-\5\ell}{2K}$	 	& $\frac{K-\5\ell}{K}$\\ \hline

 $\varphi^{\5\ell,-1}_{m}$			& $1$		& $+$		& $\5\ell\in 2\Z, \quad 0\leq \5\ell \leq k$ 	
& $\begin{array}{c} m\notin 2\Z, \quad |m| < \5\ell\\ (|m|-1)^2 \geq \5\ell(\5\ell+2)+1-K \end{array}$ 	& $(K-1)(\5\ell-m-1)-1-\b K$	& $\frac{3K+2}{8K}-h^{\5\ell,-1}_{m}-\frac{1}{2} q^{\5\ell,-1}_{m}$	& $\frac{K+2}{2K}-q^{\5\ell,-1}_{m}$\\ \hline
						& $3$		& $+$		& $\5\ell\notin 2\Z, \quad 0\leq \5\ell \leq k$
& $\begin{array}{c} m\in 2\Z, \quad |m| < \5\ell\\ (|m|-1)^2 \geq \5\ell(\5\ell+2)+1-K \end{array}$		& $-(K+1)(\5\ell-m+1)+1-\b K$	& $\frac{3K+2}{8K}-h^{\5\ell,-1}_{m}-\frac{1}{2} q^{\5\ell,-1}_{m}$	&$\frac{K+2}{2K}-q^{\5\ell,-1}_{m}$\\   \hline

 $\varphi^{\5\ell,1}_{m}$			& $1$		& $-$		& $\5\ell\notin 2\Z, \quad 0\leq \5\ell \leq k$ 		
& $\begin{array}{c} m\in 2\Z, \quad |m\pm K| \le K-2-\5\ell\\ (|m\pm K|-1)^2 \geq (K-2-\5\ell)(K-\5\ell)+1-K \end{array}$ 	
					& $(K-1)(\5\ell-m+1)+1-\b K$	& $\frac{3K+2}{8K}-h^{K-2-\5\ell,-1}_{m\pm K}-\frac{1}{2} q^{K-2-\5\ell,-1}_{m\pm K}$	& $\frac{K+2}{2K}-q^{K-2-\5\ell,-1}_{m\pm K}$\\   
\hline
						& $3$		& $-$		& $\5\ell\in 2\Z, \quad 0\leq \5\ell \leq kpage$
& $\begin{array}{c} m\notin 2\Z, \quad |m\pm K| \le K-2-\5\ell\\ (|m\pm K|-1)^2 \geq (K-2-\5\ell)(K-\5\ell)+1-K \end{array}$  			& $(K+1)(-\5\ell+m+1)-1-\b K$	& $\frac{3K+2}{8K}-h^{K-2-\5\ell,-1}_{m\pm K}-\frac{1}{2} q^{K-2-\5\ell,-1}_{m\pm K}$	& $\frac{K+2}{2K}-q^{K-2-\5\ell,-1}_{m\pm K}$ \\\hline \hline
 \end{tabular}}}
\caption{Left-moving $\C'$-sector charges $q_R'$ and conformal weights $h_R'$ (in the Ramond sector), signs $\sigma'=(-1)^{\g}$ and constraints for the admissible terms in the CPP in $\C'$ ~$\sim \sigma' x^{\nu/2} q^{h_R'-\frac{c'}{24}} t^{q_R'}$.}
\label{NAgen}
\end{center}
\end{sidewaystable}

\newpage
\subsection{Counting antigenerations}

In complete analogy to the counting of generations we have to depict a right-moving representative and compute admissible left-moving states on its orbit. We choose the same space-time matter scalar as in (\ref{sec:countingGEN}).
In order to count antigenerations, as represented by $\bf{\ovl{16}}$ in (\ref{BW}), the left-moving states must transform under $SO(8)\times U(1)$ as  $\mathbf{8}^{v}_{1}$ or $\mathbf{8}^{\5s}_{-1}$. Again, for convenience, we stay in the NS sector where the admissible states are of the general form 
\be
\Psi_{\rm left} = |\Phi \otimes \varphi \,;\, D_1 \otimes V\rangle_{l}.
\ee
With the condition for the conformal weights and charges of massless states we get four possibilities for admissible left-moving states. Their space-time parts, conformal weights and $U(1)$ charges are   
\be
{\renewcommand{\arraystretch}{2.2}
\renewcommand{\tabcolsep}{0.08cm}
\begin{array}{lll}
\label{allleftanti}
&|\id \otimes V\rangle_{l} \quad \text{with} \quad h_\intern=\frac12\, ,\;q_\intern=1; \qquad
&|v \otimes V\rangle_{l} \quad \text{with} \quad h_\intern=0 \, ,\; q_\intern=\frac12\\
&|s \otimes V\rangle_{l} \quad \text{with} \quad h_\intern=\frac38 \, ,\; q_\intern=\frac12; \qquad
&|\5s \otimes V\rangle_{l} \quad \text{with} \quad h_\intern=\frac38 \, ,\; q_\intern=\frac32.
\end{array}}
\ee

\subsubsection{Aligned antigenerations}

For an even power of the Bonn twist there is alignment within the space-time part. Taking into account the BPS bound the only states that are admissible have are of the form
\be
\Psi_{\rm left} = |\Phi \otimes \varphi^{\ell,0}_{-\ell}\,;\, \id \otimes V\ket_l	\qquad \text{with} \quad h_\intern=\frac12\,,\; q_\intern=1
\ee
with $\ell=0,\dots,K-2$. Internal states are chiral primary states. The $U(1)$ charge of the chiral primary state $\varphi^{\ell,0}_{-\ell}$ can easily be computed, and the charge contributions from the $\F$ and $\C'$ sectors to $q_\intern=1$, hence, are
\be
q^F_{\CP}=\frac{\ell}{K}	\qquad \text{and}	\qquad	q'_{\CP}=\frac{K-\ell}{K}.
\ee
A similar analysis as in the case of aligned generations can be carried out to yield admissible terms in the EPP of $\C'$ which are summarized in table \ref{Aantigen}.
\begin{table}[!h]
\begin{center}
{\renewcommand{\arraystretch}{1.5}
\renewcommand{\tabcolsep}{0.3cm}
{\small \begin{tabular}{||c|c|c|c|c||} \hline \hline
\multicolumn5{||c||}{\bf $\ovl{16}$ - aligned antigenerations}\\ \hline \hline
$\sigma'$ 	& $\5\ell$						& l	 														& $\5q'_{\CP}$ 			& $q'_{\CP}$ \\ \hline \hline
$+$		& $\5\ell \in 2\Z, \quad  0\leq \5\ell \leq k$		& $\begin{array}{ll} l\in 2K\Z+1 \qquad &\b=0, \; n \in 2\Z\\  l\in2K\Z \qquad &\b=2, \; n \notin 2\Z \end{array}$	& $\frac{K+2+\5\ell}{K}$	& $\frac{K-\5\ell}{K}$\\ \hline
$-$		& $\5\ell \notin 2\Z,  \quad 0\leq \5\ell \leq k$	& $\begin{array}{ll} l\in 2K\Z \quad &\b=0, \; n \notin 2\Z\\  l\in 2K\Z+1 \qquad &\b=2, \; n \in 2\Z \end{array}$	& $\frac{K+2+\5\ell}{K}$	& $\frac{\5\ell+2}{K}$\\
\hline \hline
 \end{tabular}}}
\caption{Left- and right-moving $\C'$-sector charges $q'_{\CP}$ and $\5q'_{\CP}$, right-moving label $\5\ell$, exponent $l$ of $x$ and sign $\sigma'=(-1)^{\g}$ in the EPP of $\C'$ with terms $~\sim \sigma' \ x^l\ t^{q'}\ \5t^{\5q'}$.}
\label{Aantigen}
\end{center}
\end{table}

\subsubsection{Non-aligned antigenerations}

For an odd power of the Bonn twist the alignment in the internal sector is broken. States in $\C'$ are in the NS sector, while states in $\F$ are in the Ramond sector. Repeating the same analysis as for non-aligned generations the only admissible left-moving states turn out to be of the form
\be
\Psi_{\rm left} = |\Phi \otimes \varphi \,;\, s \otimes V\ket_l	\qquad \text{with}	\quad h_\intern=\frac{3}{8}\,,\; q_\intern=\frac12.
\ee
Admissible states in the $\F$ sector can be derived along the same lines as for non-aligned antigenerations. Using the information about their conformal weights and $U(1)$ charges admissible terms in the {complementary Poincar\'{e}} polynomial $\C'$ can be determine and are listed in table \ref{NAantigen}.

\begin{sidewaystable}
\begin{center}
{\renewcommand{\arraystretch}{2.2}
\renewcommand{\tabcolsep}{0.08cm}
{\scriptsize \begin{tabular}{||c|c|c|c|c|c|c|c||} \hline \hline
 \multicolumn8{||c||}{\bf $\ovl{16}$ - non-aligned antigenerations}\\ \hline \hline
Fermat						& $K \mod 4$ 	& $\sigma'$ 	& $\5\ell$				
&$m$			& $\nu \mod 4K$			& $h_R'$ 			& $q_R'$\\ \hline \hline

 $\varphi^{\5\ell,1}_{\5\ell+1}$		& $1$		& $+$		& $\5\ell \in 2\Z, \quad 0\leq \5\ell \leq \frac{K-1}{2}$	
& $m=\5\ell+1$ 		& $1-\b K$			& $\frac{\5\ell+2+K}{2K}$		& $\frac{\5\ell+2+K}{K}$\\ \hline
						& $3$		& $+$		& $\5\ell\notin 2\Z, \quad 0\leq \5\ell \leq \frac{K-1}{2}$
& $m=\5\ell+1$ 		& $1-\b K+2K$		 	& $\frac{\5\ell+2+K}{2K}$		& $\frac{\5\ell+2+K}{K}$ \\\hline

 $\varphi^{\5\ell, -1}_{-(\5\ell+1)}$		& $1$		& $-$		& $\5\ell\notin 2\Z, \quad \frac{K-3}{2} \leq \5\ell \leq k$
& $m=-(\5\ell+1)$	& $-1-2\5\ell-\b K+2K$		& $\frac{2K-\5\ell}{2K}$		& $\frac{2K-\5\ell}{K}$\\ \hline
						& $3$		& $-$		& $\5\ell \in 2\Z, \quad \frac{K-3}{2} \leq \5\ell \leq k$ 	
& $m=-(\5\ell+1)$	& $-1-2\5\ell-\b K+2K$		& $\frac{2K-\5\ell}{2K}$	 	& $\frac{2K-\5\ell}{K}$\\ \hline

 $\varphi^{\5\ell,-1}_{m}$			& $1$		& $+$		& $\5\ell\in 2\Z, \quad 0\leq \5\ell \leq k$ 	
& $\begin{array}{c} m\notin 2\Z, \quad |m| < \5\ell\\ (|m|-1)^2 \geq \5\ell(\5\ell+2)+1-K \end{array}$ 	& $(K-1)(\5\ell-m+1)+1-\b K$	& $\frac{7K+2}{8K}-h^{\5\ell,1}_{m}-\frac{1}{2} q^{\5\ell,1}_{m}$	& $\frac{3K+2}{2K}-q^{\5\ell,1}_{m}$\\ \hline
						& $3$		& $+$		& $\5\ell\notin 2\Z, \quad 0\leq \5\ell \leq k$
& $\begin{array}{c} m\in 2\Z, \quad |m| < \5\ell\\ (|m|-1)^2 \geq \5\ell(\5\ell+2)+1-K \end{array}$		& $-(K+1)(\5\ell-m+1)-1-\b K$	& $\frac{7K+2}{8K}-h^{\5\ell,1}_{m}-\frac{1}{2} q^{\5\ell,1}_{m}$	&$\frac{3K+2}{2K}-q^{\5\ell,1}_{m}$\\   \hline

 $\varphi^{\5\ell,1}_{m}$			& $1$		& $-$		& $\5\ell\notin 2\Z, \quad 0\leq \5\ell \leq k$ 		
& $\begin{array}{c} m\in 2\Z, \quad |m\pm K| \le K-2-\5\ell\\ (|m\pm K|-1)^2 \geq (K-2-\5\ell)(K-\5\ell)+1-K \end{array}$ 	
					& $(K-1)(\5\ell-m-1)-1-\b K$	& $\frac{7K+2}{8K}-h^{K-2-\5\ell,1}_{m\pm K}-\frac{1}{2} q^{K-2-\5\ell,1}_{m\pm K}$	& $\frac{3K+2}{2K}-q^{K-2-\5\ell,1}_{m\pm K}$\\   
\hline
						& $3$		& $-$		& $\5\ell\in 2\Z, \quad 0\leq \5\ell \leq k$
& $\begin{array}{c} m\notin 2\Z, \quad |m\pm K| \le K-2-\5\ell\\ (|m\pm K|-1)^2 \geq (K-2-\5\ell)(K-\5\ell)+1-K \end{array}$  			& $(K+1)(-\5\ell+m-1)+1-\b K$	& $\frac{7K+2}{8K}-h^{K-2-\5\ell,1}_{m\pm K}-\frac{1}{2} q^{K-2-\5\ell,1}_{m\pm K}$	& $\frac{3K+2}{2K}-q^{K-2-\5\ell,1}_{m\pm K}$ \\\hline \hline
 \end{tabular}}}
\caption{Left-moving $\C'$-sector charges $q_R'$ and conformal weights $h_R'$ (in the Ramond sector), signs $\sigma'=(-1)^{\g}$ and constraints for the admissible terms in the CPP in $\C'$ ~$\sim \sigma' x^{\nu/2} q^{h_R'-\frac{c'}{24}} t^{q_R'}$.}
\label{NAantigen}
\end{center}
\end{sidewaystable}

\subsection{Counting vectors}

As before, we choose again the space-time matter scalar as right-moving representative. In order to count vectors, as represented by $\bf{10}$ in (\ref{BW}), left-moving states must transform under $SO(8) \times U(1)$ as ${\bf1}_{2}$, ${\bf1}_{-2}$ or ${\bf8}^s_0$. Let us consider states transforming under ${\bf1}_{2}$ the general form of which is given by
\be
\Psi_{\rm left}=|\Phi \otimes \varphi \,;\, D_1 \otimes \id\ket_l.
\ee
The space-time parts, conformal weights and $U(1)$ charges of admissible states contributing to the massless spectrum are
\be
{\renewcommand{\arraystretch}{2.2}
\renewcommand{\tabcolsep}{0.08cm}
\begin{array}{lll}
\label{allleftvec}
&|\id \otimes \id \rangle_{l} \quad \text{with} \quad h_\intern=1\, ,\;q_\intern=2; \qquad
&|v \otimes \id \rangle_{l} \quad \text{with} \quad h_\intern=\frac12 \, ,\; q_\intern=1\\
&|s \otimes \id \rangle_{l} \quad \text{with} \quad h_\intern=\frac78 \, ,\; q_\intern=\frac32; \qquad
&|\5s \otimes \id \rangle_{l} \quad \text{with} \quad h_\intern=\frac78 \, ,\; q_\intern=\frac52.
\end{array}}
\ee
States with $q_\intern=\frac52$ can already be discarded since they do not obey the BPS bound.

\subsubsection{Aligned vectors}

For an even power of the Bonn twist there are now two possible states, both of which have internal chiral primary states.
States with $q_\intern=2$ are of the form
\be
\Psi_{\rm left}=|\Phi \otimes \varphi^{\ell,0}_{-\ell} \,;\, \id \otimes \id\ket_l	\qquad \quad \text{with} \quad q^F_{\CP}=\frac{\ell}{K}	\quad \text{and}	\quad q'_{\CP}=\frac{2K-\ell}{K},
\ee
while states with $q_\intern=1$ are of the form
\be
\Psi_{\rm left}=|\Phi \otimes \varphi^{-\ell,0}_{\ell} \,;\, v \otimes \id\ket_l 	\qquad \quad \text{with} \quad q^F_{\CP}=\frac{\ell}{K}	\quad \text{and}	\quad q'_{\CP}=\frac{K-\ell}{K},
\ee
and $\ell=0,\dots,k$.
Admissible terms in the EPP of $\C'$ are listed in table \ref{Avec}.
\begin{table}[!h]
\begin{center}
{\renewcommand{\arraystretch}{1.5}
\renewcommand{\tabcolsep}{0.3cm}
{\small \begin{tabular}{||c|c|c|c|c|c||} \hline \hline
\multicolumn6{||c||}{\bf $10$ - aligned vectors}\\ \hline \hline
$q_{int}$	&$\sigma'$ 	& $\5\ell$						& l	 														& $\5q'_{\CP}$ 			& $q'_{\CP}$ \\ \hline \hline
$2$		&$-$		& $\5\ell \in 2\Z,  \quad 0\leq \5\ell \leq k$	& $\begin{array}{ll} l\in 2K\Z+1 \qquad &\b=0, \; n \in 2\Z\\  l\in 2K\Z \qquad &\b=2, \; n \notin 2\Z \end{array}$	& $\frac{K+2+\5\ell}{K}$	& $\frac{2K-\5\ell}{K}$\\
\hline 
$2$		&$+$		& $\5\ell \notin 2\Z, \quad  0\leq \5\ell \leq k$		& $\begin{array}{ll} l\in 2K\Z \qquad &\b=0, \; n \notin 2\Z\\  l\in2K\Z+1 \qquad &\b=2, \; n \in 2\Z \end{array}$	& $\frac{K+2+\5\ell}{K}$	& $\frac{K+2+\5\ell}{K}$\\ \hline \hline
$1$		&$+$		& $\5\ell \notin 2\Z, \quad  0\leq \5\ell \leq k$& $\begin{array}{ll} l\in 2K\Z+1 \qquad &\b=0, \; n \in 2\Z\\  l\in2K\Z \qquad &\b=2, \; n \notin 2\Z \end{array}$		& $\frac{K+2+\5\ell}{K}$	& $\frac{K-\5\ell}{K}$\\ \hline
$1$		&$-$		& $\5\ell \in 2\Z,  \quad 0\leq \5\ell \leq k$	& $\begin{array}{ll} l\in 2K\Z \qquad &\b=0, \; n \notin 2\Z\\  l\in 2K\Z+1 \qquad &\b=2, \; n \in 2\Z \end{array}$			& $\frac{K+2+\5\ell}{K}$	& $\frac{\5\ell+2}{K}$\\
\hline \hline
 \end{tabular}}}
\caption{Left- and right-moving $\C'$-sector charges $q'_{\CP}$ and $\5q'_{\CP}$, right-moving label $\5\ell$, exponent $l$ of $x$ and sign $\sigma'=(-1)^{\g}$ in the EPP of $\C'$ with terms $~\sim \sigma' \ x^l\ t^{q'}\ \5t^{\5q'}$.}
\label{Avec}
\end{center}
\end{table}

\subsubsection{Non-aligned vectors}

Due to the BPS bound there are only states of the form
\be
\Psi_{\rm left}=|\Phi \otimes \phi \,;\, \5s \otimes \id\ket_l		\qquad \text{with} \quad h_\intern=\frac78\,;\; q_\intern=\frac32
\ee
that can contribute to the massless spectrum. Again, states in $\C'$ are in the NS sector while states in $\F$ are in the Ramond sector. Those can either be Ramond ground states or excited states, as discussed already for non-aligned generations. The complete list of constraints that have to be satisfied by states in $C'$ in order to yield admissible terms in the {complementary Poincar\'{e}} polynomial together with their conformal weights and $U(1)$ charges is given in table \ref{NAvec}.

\begin{sidewaystable}
\begin{center}
{\renewcommand{\arraystretch}{2.2}
\renewcommand{\tabcolsep}{0.08cm}
{\scriptsize \begin{tabular}{||c|c|c|c|c|c|c|c||} \hline \hline
 \multicolumn8{||c||}{\bf 10 - non-aligned vectors}\\ \hline \hline
Fermat						& $K \mod 4$ 	& $\sigma'$ 	& $\5\ell$				
&$m$			& $\nu \mod 4K$			& $h_R'$ 			& $q_R'$\\ \hline \hline

 $\varphi^{\5\ell,1}_{\5\ell+1}$		& $1$		& $-$		& $\5\ell \in 2\Z, \quad 0\leq \5\ell \leq \frac{K-1}{2}$	
& $m=\5\ell+1$ 		& $1-\b K$			& $\frac{\5\ell+2+3K}{2K}$		& $\frac{\5\ell+2+2K}{K}$\\ \hline
						& $3$		& $-$		& $\5\ell\notin 2\Z, \quad 0\leq \5\ell \leq \frac{K-1}{2}$
& $m=\5\ell+1$ 		& $1-\b K+2K$		 	& $\frac{\5\ell+2+3K}{2K}$		& $\frac{\5\ell+2+2K}{K}$ \\\hline

 $\varphi^{\5\ell, -1}_{-(\5\ell+1)}$		& $1$		& $+$		& $\5\ell\notin 2\Z, \quad \frac{K-3}{2} \leq \5\ell \leq k$
& $m=-(\5\ell+1)$	& $-1-2\5\ell-\b K+2K$		& $\frac{4K-\5\ell}{2K}$		& $\frac{3K-\5\ell}{K}$\\ \hline
						& $3$		& $+$		& $\5\ell \in 2\Z, \quad \frac{K-3}{2} \leq \5\ell \leq k$ 	
& $m=-(\5\ell+1)$	& $-1-2\5\ell-\b K+2K$		& $\frac{4K-\5\ell}{2K}$	 	& $\frac{3K-\5\ell}{K}$\\ \hline

 $\varphi^{\5\ell,-1}_{m}$			& $1$		& $-$		& $\5\ell\in 2\Z, \quad 0\leq \5\ell \leq k$ 	
& $\begin{array}{c} m\notin 2\Z, \quad |m| < \5\ell\\ (m-1)^2 \geq \5\ell(\5\ell+2)+1-K \end{array}$ 	& $(K-1)(\5\ell-m+1)+1-\b K$	& $\frac{15K+2}{8K}-h^{\5\ell,1}_{m}-\frac{1}{2} q^{\5\ell,1}_{m}$	& $\frac{5K+2}{2K}-q^{\5\ell,1}_{m}$\\ \hline
						& $3$		& $-$		& $\5\ell\notin 2\Z, \quad 0\leq \5\ell \leq k$
& $\begin{array}{c} m\in 2\Z, \quad |m| < \5\ell\\ (|m|-1)^2 \geq \5\ell(\5\ell+2)+1-K \end{array}$		& $-(K+1)(\5\ell-m+1)-1-\b K$	& $\frac{15K+2}{8K}-h^{\5\ell,1}_{m}-\frac{1}{2} q^{\5\ell,1}_{m}$	&$\frac{5K+2}{2K}-q^{\5\ell,1}_{m}$\\   \hline

 $\varphi^{\5\ell,1}_{m}$			& $1$		& $+$		& $\5\ell\notin 2\Z, \quad 0\leq \5\ell \leq k$ 		
& $\begin{array}{c} m\in 2\Z, \quad |m\pm K| \le K-2-\5\ell\\ (m\pm K-1)^2 \geq (K-2-\5\ell)(K-\5\ell)+1-K \end{array}$ 	
					& $(K-1)(\5\ell-m-1)-1-\b K$	& $\frac{15K+2}{8K}-h^{K-2-\5\ell,1}_{m\pm K}-\frac{1}{2} q^{K-2-\5\ell,1}_{m\pm K}$	& $\frac{5K+2}{2K}-q^{K-2-\5\ell,1}_{m\pm K}$\\   
\hline
						& $3$		& $+$		& $\5\ell\in 2\Z, \quad 0\leq \5\ell \leq k$
& $\begin{array}{c} m\notin 2\Z, \quad |m\pm K| \le K-2-\5\ell\\ (|m\pm K|-1)^2 \geq (K-2-\5\ell)(K-\5\ell)+1-K \end{array}$  			& $(K+1)(-\5\ell+m-1)+1-\b K$	& $\frac{15K+2}{8K}-h^{K-2-\5\ell,1}_{m\pm K}-\frac{1}{2} q^{K-2-\5\ell,1}_{m\pm K}$	& $\frac{5K+2}{2K}-q^{K-2-\5\ell,1}_{m\pm K}$ \\\hline \hline
 \end{tabular}}}
\caption{Left-moving $\C'$-sector charges $q_R'$ and conformal weights $h_R'$ (in the Ramond sector), signs $\sigma'=(-1)^{\g}$ and constraints for the admissible terms in the CPP in $\C'$ ~$\sim \sigma' x^{\nu/2} q^{h_R'-\frac{c'}{24}} t^{q_R'}$.}
\label{NAvec}
\end{center}
\end{sidewaystable}

\newpage

\section{Distler-Kachru models and \\the heterotic (0,2) CFT/geometry connection}

In analogy to the case of (2,2) models, a very general framework for the description of (0,2) models can be given in terms of a gauged linear sigma model with $(0,2)$ worldsheet supersymmetry, known as Dister-Kachru models. Since we want to compare the spectra obtained by the counting algorithm of the previous chapter to that of Dister-Kachru models\cite{di94} let us briefly review their structure.
In (0,2) models there exists an additional structure, as compared to (2,2) models, which is the choice of rank $\widetilde{r}$ stable, holomorphic vector bundle $V\to \M$ with vanishing first Chern class $c_1(V)=0$ and $c_2(V)=c_2(T)$, where $T$ is the holomorphic tangent bundle of $\M$. 
As reviewed in \cite{bl96}
, the defining data of a (0,2) sigma model on a Calabi-Yau manifold $\M$ is encoded in the superpotentials $W_j(\Phi_i)$ and $F^l_a(\Phi_i)$, where $W_j(\Phi_i)$ are transversal polynomials of degree $d_j$ which define the base space $\M$ of the vector bundle $V \to \M$ associated to the left-moving gauge fermions and $F^l_a(\Phi_i)$ are polynomials, with degree fixed by requiring charge neutrality of the action, that define the global structure of the bundle $V$.
The field content is given by a set of chiral superfields $\Phi_i$ with $U(1)$ charges $w_i$ with $i=1,\cdots,N_i$. Neutrality of the action then requires additional Fermi superfields $\Sigma^j$ with charge $-d_j$ with $j=1,\cdots,N_j$. The ingredients for constructing the bundle $V$ are Fermi superfields $\Lambda^a$ with strictly positive~\footnote{If the $n_a$ are not strictly positive, the bundle $V$ is never stable.\cite{di94}} $U(1)$ charges $n_a$ with $a=1,\cdots,N_a$ and a chiral superfield $P_l$ with charge $-m_l$ with $l=1,\cdots,N_l$ such that $\sum_l m_l=\sum_a n_a$. The (0,2) superpotential action that summarizes the structure of the total bundle is given by
\be
S_{\cal W}=\int d^2z d\theta \Big(\Sigma^j W_j(\Phi_i) + P_l \Lambda^a F^l_a(\Phi_i)\Big)\ .
\ee
The first term ensures that the fields $\Phi_i$ lie on the hypersurface $W_j=0$, whereas the second term ensures that the gauge fermions $\lambda^a$ (lowest components of the $\Lambda^a$) are sections of the bundle $V$.
The (0,2) gauge multiplets are determined by a real superfield ${\cal V}$, which contains the right-moving component of the gauge field, and a superfield ${\cal A}$, which contains the left-moving component of the gauge field.

The structure of the vector bundle $V$ of rank $\widetilde{r}=N_a-N_l$ is given by the short exact sequence (monad)                          
\BE                                                           
        0~\rightarrow~ V ~\rightarrow~ \bigoplus^{\widetilde{r}+N_l}_{a=1}{\cal O}(n_{a})
        \stackrel{F_a}{~\longrightarrow~} \bigoplus^{N_l}_{l=1}{\cal O}(m_l)~\rightarrow~ 0
\label{monad}
\EE
with Chern class
\be
c(V)=\frac{c\big(\oplus_{a=1}^{\widetilde{r}+N_l}{\cal O}(n_a)\big)}{c\big(\oplus_{l=1}^{N_l}{\cal O}(m_l)\big)}\ .
\ee
Restricting to the case of $N_l=1$ the exact sequence defines a vector bundle of rank $\widetilde{r}=N_a-1$ over a complete intersection Calabi-Yau variety $\M$. The $F_a$ are homogeneous polynomials of degrees $m-n_a$ which do not vanish simultaneously on $\M$. For weighted projective ambient spaces we can write this data as
\BE                                                    
        V_{n_1\ldots,n_{\widetilde{r}+1}}[m]~\longrightarrow
                ~{\Bbb P}_{w_{1},\ldots , w_{N_j+4}}[d_1,\ldots,d_{N_j}]\ ,
\EE 
where $N_j$ is the codimension of the Calabi-Yau manifold and $\widetilde{r}=4,5$ corresponds to unbroken gauge groups $SO(10)$ and $SU(5)$, respectively.
\\
The Calabi-Yau condition $c_1(T)=0$ and the condition $c_{1}(V)=0$ imply
\begin{equation}                                               
                   \sum_j d_j-\sum_i w_i= m-\sum_a n_{a}=0 \label{VBc1}\ .
\end{equation}
Cancellation of gauge anomalies
$ch_{2}(V)=ch_{2}(TX)$ with the second Chern character $ch_{2}=\frac{1}{2}c_{1}^{2}-c_{2}$ implies 
the quadratic Diophantine constraint                           
\begin{equation}                                             
        \sum_j
        d_{j}^{2} -\sum_i
                w_{i}^{2}       =       m^{2}-\sum_a
               n_{a}^{2}\ .\label{VBc2}
\end{equation}
In general there are not many solutions to this equation.
In the $(2,2)$ case, which corresponds to $F_a=\partial_a W$ and yields the gauge group $E_6$, a solution is given by the choice of $m=d=\sum_j w_j$ with $n_a=w_a$. \\
Note, that the discrete gauge symmetry $\Z_m$ that survives the breaking of the $U(1)$ in the gauged linear sigma model with $m$ defined in (\ref{monad}) corresponds to the $\Z_m$ quantum symmetry\cite{Vafa:1989xc} resulting from the GSO projection on the CFT side.

In \cite{bl95} 
R.~Blumenhagen and A.~Wi{\ss}kirchen proposed a Gepner-type construction of string models with $(0,2)$ worldsheet supersymmetry based on the simple current construction to obtain heterotic compactifications yielding different gauge groups and massless spectra. In \cite{bl96} 
they, together with R.~Schimmrigk, describe the analog of the (2,2) triality between exactly solvable conformal field theories, $(0,2)$ Calabi-Yau manifolds and Landau-Ginzburg theories.
The suggested CFT/geometry correspondence\cite{bl95,bl96} 
, in particular, associates the vector bundle $V_{1,1,1,1,1}[5]$ over the complete intersection Calabi-Yau $\IP_{1,1,1,1,2,2}[4,4]$ to a (0,2) cousin of the exactly solvable (2,2) Gepner model~$3^5$, which is described by the Landau-Ginzburg model $\IP_{1,1,1,1,1}[5]$ and corresponds, in the sigma model language, to the quintic Calabi-Yau manifold. Note, that the codimension of the Calabi-Yau manifold for the $(0,2)$ cousin has increased as compared to the $(2,2)$ case. The bundle data of the $(0,2)$ quintic cousin can be expressed by the exact sequence
\be
0 \to \bigoplus_{a=1}^5 \mathcal{O}(1) \to \mathcal{O}(5) \to 0 \ .
\ee
The underlying conformal field theory builds on a tensor product of five minimal model factors and a supersymmetry breaking simple current that acts only on one factor. For this class of $(0,2)$ models the Gepner model data directly determines the vector bundle structure. Since the twist, that defines the $(0,2)$ model, only acts on one of the minimal model factors, one might be tempted to expect that the conjecture can be generalized to a larger picture, where a more general form of an exactly solvable theory directly translates into the bundle data $ V_{n_1,\ldots,n_5}[m]$. In \cite{kr96,wkm} 
an ansatz for a solution to \ref{VBc1} and \ref{VBc2} was made by setting $w_i=n_i$ for $i<5$ and $w_5=2n_5$
\BE                                                        
        V_{n_1,\ldots,n_5}[m]   \to
       {\Bbb P}_{n_1,\ldots ,n_4,2n_5,w_6}[d_1,d_2]\ ,
\EE
and imposing (\ref{VBc1}) and (\ref{VBc2}) yielding                               
\BE                                                 
        d_1+d_2=m+n_5+w_6\qquad \text{and} \qquad d_1^2+d_2^2=m^2+3n_5^2+w_6^2\ .
\EE
It is quite non-trivial and encouraging that this non-linear system has a
general solution 
\be
w_6=(m-n_5)/2=d_1/2 \qquad \text{and} \qquad d_2=(m+3n_5)/2\ .
\ee
By replacing all minimal model factors of the internal conformal field theory, except the one on which the twist acts, by an arbitrary CFT the $(0,2)$ CFT/geometry correspondence needs to be adapted to generic Landau-Ginzburg models.
In \cite{kr96,wkm} 
it was conjectured that there is a non-rational extension of the $(0,2)$ CFT/geometry correspondence between the $(0,2)$ Gepner-type models and the Dister-Kachru models defined by the data
\BE                                                           
        V_{n_1,\ldots,n_5}[m]   \to                     \label{conj2}
       {\Bbb P}_{n_1,\ldots ,n_4,2n_5,\frac{m-n_5}{2}}[m-n_5,(m+3n_5)/2]\ ,
\EE
where $m/n_5$ is an odd integer and there exists a transversal polynomial
$p(z_1,\ldots,z_4)$ of degree $m$ that is quasi-homogeneous with weights
$w(z_i)=n_i$ for $i\le4$. The increase of the codimension of the Calabi-Yau manifold may be interpreted as providing an additional field of degree $w_6=d_1/2$ that generates the twisted sectors for the $\Z_2$ orbifolding due to $J_b$.

In order to test the extension of the $(0,2)$ CFT/geometry correspondence to the non-rational realm
we have to compare the spectra we obtain using the counting algorithm on the CFT side to that of non-linear sigma models at the infrared fixed point which are described by Landau-Ginzburg orbifold models.
In particular, we compare the number of generations and antigenerations as arising from both, the CFT and the geometry computations. 
For a generic choice of data in a Dister-Kachru model, defined by the stable bundle
\be
V_{n_{1}, \dots, n_{\widetilde r+1}}[m]\to \mathbb{P}_{w_{1}, \dots, w_{N_{i}}}[d_{1}, \dots, d_{N_{c}}]
\ee
of rank $\widetilde r$ over a complete intersection space of codimension $N_c$, this can be computed by using the elliptic genus $Z_{{LG}}$ as explained in \cite{bsw96,Kawai:1994qy}. Its contribution in the $\a$-th twisted sector is given by
\be
Z_{{LG}}^{\a}=Tr_{\H_{\a}}(-1)^F t^{J_0} q^{H} \sim \chi_{\a}+O(q)\ .
\ee
The $\chi$-genus of a bundle of rank $\widetilde r$ can be written as
\be
\label{eq:LG-genus}
\chi_{\alpha} = \left.
\frac{\prod_{a}(-1)^{[\alpha\,\nu_{a}]}(t^{\nu_{a}}\,q^{\beta_{a}/2})^{\{\alpha\nu_{a}\}}
(1-t^{\nu_{a}}\,q^{\{\alpha\nu_{a}\}})(1-t^{-\nu_{a}}\,q^{-\beta_{a}})}
{\prod_{i}(-1)^{[\alpha\,q_{i}]} (t^{q_{i}}\,q^{\beta_{i}/2})^{\{\alpha q_{i}\}}
(1-t^{q_{i}}\,q^{\{\alpha q_{i}\}})(1-t^{-q_{i}}\,q^{-\beta_{i}})
}
\right |_{q^{0}\,t^{n}}\ ,
\ee
where $(\cdots)|_{q^{0}\,t^{n}}$ denotes the evaluation of the $q^{0}\,t^{n}$ terms in the Laurent expansion with integer $n$ and
\be
\{x\} := x-[x],\qquad \beta_{a} := \{\alpha\,\nu_{a}\}-1,\qquad \beta_{i} := \{\alpha\,q_{i}\}-1\ .
\ee
The charges of the fields are
\be
q_{i} = \frac{w_{i}}{m},\qquad \nu_{a} = 1-\frac{n_{a}}{m}\qquad \text{and} \qquad \nu_{\widetilde r+1+l} = \frac{d_{l}}{m}\ . 
\ee
The number of generations is the sum of the positive coefficients of monomials in $t^{1}$
(as $\a$ varies), while the number of antigenerations is the sum of positive coefficients of monomials in $t^{3}$. \\
These numbers are independent on the defining DK data\footnote{By this, we mean that the $\chi$-genus does not depend on the form of the superpotentials $W$ or $F$ which are the defining data of a specific DK model. Nevertheless, we will call these $(0,2)$ LG models Distler-Kachru models in order to emphasize that they have a geometric and a CFT phase.} and are reliable if no extra gauginos or generation/antigeneration pairings occur. In the latter case it turns out that the number of generations $n_{\bf N}$ and antigenerations $n_{\bf \5N}$ need not be constant over the moduli space. In any case, the number of net generations $n_{net}=n_{\bf{N}}-n_{\bf{\5{N}}}$ is constant in moduli space as it is given by an index theorem $n_{net}=\Big|\frac12 \int c_3(V)\Big|$. \cite{di94,bsw96} 
For vectors it is more subtle. Since mass terms for states transforming in ${\bf 10}$ are not forbidden in the spacetime superpotential, the number of vectors might jump as we move from the Calabi-Yau phase to the Landau-Ginzburg phase.\cite{di94} 

Further subtleties arise when extra massless gauginos occur in the spectrum which, in \cite{di94}, is described to be the analog of the destabilization of the vacuum by worldsheet instantons in the Calabi-Yau phase. In this case the DK model might be sensitive to generic choices of its defining data and only certain constraints might lead to ``honest'' $(0,2)$ SCFTs in the infrared limit. However, these might not have the desired gauge group. For further reference see, in particular, \cite{di94,bsw96}.

\section{Checks and examples}

So far the conjectured CFT/geometry correspondence is only based on the existence of a ``natural'' 
solution to the anomaly cancellation constraints. We can test it by 
working out the spectra by two different methods.
\begin{enumerate}
\item On the CFT side we use the counting algorithm that we have derived in the previous sections
and which works for a generic LG model. We can compute the number of generations, antigenerations and vectors.
\item On the DK side we use the elliptic genus to compute the Euler characteristic of the bundle.
If no extra gauginos or generation/antigeneration pairings occur, it is possible to extract the number of 
generations and antigenerations, separately, as explained in \cite{bsw96}. 
\end{enumerate}

In the following we will consider various examples including Fermat-type and non-Fermat-type LG models. As a prominent example of Fermat-type models we show that the number of generations, antigenerations and vectors of the (0,2) cousin of the quintic as computed on the CFT side by our counting algorithm agrees with those first calculated in \cite{bl95}. A couple of non-Fermat-type examples are shown to give the same numbers of generations and antigenerations on the CFT and the DK side. 
A couple of non-Fermat-type examples are shown to give the same numbers of generations and antigenerations on the CFT side as that computed by the $\chi$-genus of DK models. 
Counting methods for both, Fermat- and non-Fermat LG models have been computerized, hence allowing for a large class of LG models to be easily tested.

\subsection{Fermat-type LG models}
We consider the following three models of type $(k'_1,...,k'_n;k)$ with one minimal model factor $\F$ of level $k$ and $i=1,...,n$ minimal model factors of level $k'_i$ that comprise $\C'$. The results of the Blumenhagen-Wi{\ss}kirchen algorithm carried out in \cite{bl95,bl96,bsw96} 
is given in table \ref{BWtable}. 
We compute the number of generations, antigenerations and vectors by determining admissible terms in the extended Poincar\'{e} polynomial and the complementary Poincar\'{e} polynomial of $\C'$ using the data and constraints from the tables derived in the previous section. As an illustrative example the EPP and CPP of $\C'$ for the (0,2) quintic cousin are given in the appendix together with a detailed analysis of the counting of generations, antigenerations and vectors.\\
\begin{table}[!h]
\begin{center}
{\renewcommand{\arraystretch}{1.3}
\renewcommand{\tabcolsep}{0.2cm}
\begin{tabular}{c|l|l|l}
${\rm model}$ & 
$N_{\bf16} = N_{\bf16}^{\rm A} + N_{\bf16}^{\rm NA}$ & 
$N_{\bf \ovl{16}} = N_{\bf\ovl{16}}^{\rm A} + N_{\bf \ovl{16}}^{\rm NA}$ &
$N_{\bf 10} = (N_{\bf10}^{\rm A_2}+ N_{\bf10}^{\rm A_1}) + N_{\bf10}^{\rm NA}$\\
\hline
$(3,3,3,3; 3)$ & $80 = 60+20$ & $0$ & $74 = (41+1)+32$ \\
$(8,8,8; 3)$    & $113 = 85+28$ & $5 = 1+4$ & $108 = (60+0)+48$ \\
$(2, 2, 8, 3; 3)$ & $34 = 24+10$ & $10 = 8+2$ & $40 = (15+7)+18$
\end{tabular}}
\caption{Number of generations, antigenerations and vectors for the models $3^4\otimes 3$, $8^3 \otimes 3$ and $2^2 \otimes 8 \otimes 3 \otimes 3$ with the abbreviations: $A$ = aligned, $AN$ = non-aligned; $A_1$ = aligned with $q_{int}=1$, $A_2$ = aligned with $q_{int}=2$.}
\label{BWtable}
\end{center}
\end{table}

\subsubsection*{Counting in $(3,3,3,3; 3)$}

The number of aligned generations is computed by summing up the coefficients of all admissible terms in the EPP of $\C'$ which are characterized by the relevant data listed in the table below.
\begin{center}
{\renewcommand{\arraystretch}{1.5}
\renewcommand{\tabcolsep}{0.5cm}
{\small \begin{tabular}{||c|c|c|c|c||c||} \hline \hline
\multicolumn6{||c||}{\bf 16 - aligned generations} \\ \hline \hline
$\sigma'$ 	& $\5\ell$	& $l$	& $\5q'_{\CP}$ 	& $q'_{\CP}$ 	& $N_{\bf 16}^A$\\ \hline \hline
$+$		& $0$		& $0$	& $\frac75$	& $\frac75$	& ${\bf40}$\\ \hline
$+$		& $2$		& $0$	& $\frac95$	& $\frac95$	& ${\bf20}$\\ \hline \hline
 \end{tabular}}}
\end{center}
Hence, the number of aligned generations is $40+20=60$. The necessary information in order to count non-aligned generations in the {complementary Poincar\'{e}} polynomial is given by the following table.
\begin{center}
 {\renewcommand{\arraystretch}{1.5}
\renewcommand{\tabcolsep}{0.5cm}
{\small \begin{tabular}{||c|c|c|c|c|c|c||c||} \hline \hline
\multicolumn8{||c||}{\bf 16 - non-aligned generations}\\ \hline \hline
Fermat					& $K\mod4$	& $\sigma'$ 	& $\5\ell$	& $\nu \mod4K$	& $h'_{R}$ 	& $q'_{R}$ 	& $N_{\bf 16}^{NA}$\\ \hline \hline
$\varphi^{\5\ell,-1}_{-(\5\ell+1)}$	& $1$		& $+$		& $2$		& $10$		& $\frac{3}{10}$					& $\frac35$	& ${\bf 20}$\\ \hline \hline
 \end{tabular}}}
\end{center}
The number of non-aligned generations is $20$ which, together with the $60$ of the aligned generations, sums up to the famous number of $80$ generations for the $(0,2)$ cousin of the quintic. There are no antigenerations in this model. In order to count aligned vectors we need the data of the following table.
\begin{center}
{\renewcommand{\arraystretch}{1.5}
\renewcommand{\tabcolsep}{0.5cm}
{\small \begin{tabular}{||c|c|c|c|c|c||c||} \hline \hline
\multicolumn7{||c||}{\bf 10 - aligned vectors} \\ \hline \hline
$q_\intern$	& $\sigma'$ 	& $\5\ell$	& l	& $\5q'_{\CP}$ 	& $q'_{\CP}$ 	& $N_{\bf10}^A$\\ \hline \hline
2		& $+$		& $1$		& $0$	& $\frac85$	& $\frac85$	& ${\bf31}$\\ \hline
2		& $+$		& $3$		& $0$	& $\frac{10}{5}$& $\frac{10}{5}$& ${\bf10}$\\ \hline 
1		& $+$		& $1$		& $2$	& $\frac{8}{5}$	& $\frac{4}{5}$& ${\bf1}$\\ \hline \hline
 \end{tabular}}}
\end{center}
Hence, there are $42$ aligned vectors. The necessary information in order to count non-aligned vectors in the {complementary Poincar\'{e}} polynomial is contained in the following table.
\begin{center}
 {\renewcommand{\arraystretch}{1.5}
\renewcommand{\tabcolsep}{0.5cm}
{\small \begin{tabular}{||c|c|c|c|c|c|c|c||c||} \hline \hline
\multicolumn9{||c||}{\bf 10 - non-aligned vectors}\\ \hline \hline
Fermat					& $K\mod4$	& $\sigma'$ 	& $\5\ell$	& $m$	& $\nu \mod 4K$	& $h'_{R}$ 			& $q'_{R}$ 	& $N_{\bf 10}^{NA}$\\ \hline \hline
$\varphi^{\5\ell,-1}_{-(\5\ell+1)}$	& $1$		& $+$		& $1$		& $-2$	& $2$						& $\frac{19}{10}$		& $\frac{14}{5}$& ${\bf 31}$\\ \hline 
$\varphi^{\5\ell,1}_{m}$		& $1$		& $+$		& $1$		& $4$	& $8$						&  $\frac{11}{10}$		& $2$		& ${\bf 1}$\\ \hline\hline
 \end{tabular}}}
\end{center}
There are $32$ non-aligned vectors which together with the $42$ aligned vectors give a total of $74$ vectors in the $(0,2)$ cousin of the quintic.

\subsubsection*{Counting in $(8,8,8; 3)$}

We can carry out the same analyis as for the quintic cousin with the following results.

\begin{center}
{\renewcommand{\arraystretch}{1.5}
\renewcommand{\tabcolsep}{0.5cm}
{\small \begin{tabular}{||c|c|c|c|c||c||} \hline \hline
\multicolumn6{||c||}{\bf 16 - aligned generations} \\ \hline \hline
$\sigma'$ 	& $\5\ell$	& $l$	& $\5q'_{\CP}$ 	& $q'_{\CP}$ 	& $N_{\bf 16}^A$\\ \hline \hline
$+$		& $0$		& $0$	& $\frac75$	& $\frac75$	& ${\bf57}$\\ \hline
$+$		& $2$		& $0$	& $\frac95$	& $\frac95$	& ${\bf28}$\\ \hline \hline
 \end{tabular}}}
\end{center}
Hence, the number of aligned generations is $85$.
\begin{center}
 {\renewcommand{\arraystretch}{1.5}
\renewcommand{\tabcolsep}{0.5cm}
{\small \begin{tabular}{||c|c|c|c|c|c|c||c||} \hline \hline
\multicolumn8{||c||}{\bf 16 - non-aligned generations}\\ \hline \hline
Fermat					& $K\mod4$	& $\sigma'$ 	& $\5\ell$	& $\nu \mod 4K$	& $h'_{R}$ 	& $q'_{R}$ 	& $N_{\bf 16}^{NA}$\\ \hline \hline
$\varphi^{\5\ell,-1}_{-(\5\ell+1)}$	& $1$		& $+$		& $2$		& $20$				& $\frac{3}{10}$					& $\frac35$	& ${\bf 28}$\\ \hline \hline
 \end{tabular}}}
\end{center}
The number of non-aligned generations is $28$. In total there are $113$ generations. 
\begin{center}
{\renewcommand{\arraystretch}{1.5}
\renewcommand{\tabcolsep}{0.5cm}
{\small \begin{tabular}{||c|c|c|c|c||c||} \hline \hline
\multicolumn6{||c||}{\bf $\5{16}$ - aligned antigenerations}\\ \hline \hline
$\sigma'$ 	& $\5\ell$	& $l$	& $\5q'_{\CP}$ 	& $q'_{\CP}$ 	& $N_{\bf 16}^A$\\ \hline \hline
$+$		& $2$		& $3$	& $\frac35$	& $\frac95$	& ${\bf1}$\\ \hline \hline
 \end{tabular}}}
\end{center}
Hence, there is only 1 aligned antigeneration.
\begin{center}
 {\renewcommand{\arraystretch}{1.5}
\renewcommand{\tabcolsep}{0.5cm}
{\small \begin{tabular}{||c|c|c|c|c|c|c||c||} \hline \hline
\multicolumn8{||c||}{\bf $\5{16}$ - non-aligned antigenerations}\\ \hline \hline
Fermat					& $K\mod4$	& $\sigma'$ 	& $\5\ell$	& $\nu \mod 4K$	& $h'_{R}$ 		& $q'_{R}$ 	& $N_{\bf 16}^{NA}$\\ \hline \hline
$\varphi^{\5\ell,1}_{\5\ell+1}$		& $1$		& $+$		& $2$		& $16$		& $\frac{9}{10}$	& $\frac95$	& ${\bf 1}$\\ \hline
$\varphi^{-\5\ell,1}_{-(\5\ell+1)}$	& $1$		& $-$		& $3$		& $18$		& $\frac{7}{10}$	& $\frac75$	& ${\bf 3}$\\ \hline \hline
 \end{tabular}}}
\end{center}
There are $4$ non-aligned antigenerations and, hence, there are $5$ antigenerations in total.
\begin{center}
{\renewcommand{\arraystretch}{1.5}
\renewcommand{\tabcolsep}{0.5cm}
{\small \begin{tabular}{||c|c|c|c|c|c||c||} \hline \hline
\multicolumn7{||c||}{\bf 10 - aligned vectors}\\ \hline \hline
$q_\intern$	& $\sigma'$ 	& $\5\ell$	& l	& $\5q'_{\CP}$ 	& $q'_{\CP}$ 	& $N_{\bf10}^A$\\ \hline \hline
2		& $+$		& $1$		& $0$	& $\frac85$	& $\frac85$	& ${\bf45}$\\ \hline
2		& $+$		& $3$		& $0$	& $\frac{10}{5}$& $\frac{10}{5}$& ${\bf15}$\\ \hline \hline
 \end{tabular}}}
\end{center}
Hence, there are $60$ aligned vectors.
\begin{center}
 {\renewcommand{\arraystretch}{1.5}
\renewcommand{\tabcolsep}{0.5cm}
{\small \begin{tabular}{||c|c|c|c|c|c|c|c||c||} \hline \hline
\multicolumn9{||c||}{\bf 10 - non-aligned vectors}\\ \hline \hline
Fermat					& $K\mod4$	& $\sigma'$ 	& $\5\ell$	& $m$	& $\nu \mod 4K$	& $h'_{R}$ 	& $q'_{R}$ 	& $N_{\bf 10}^{NA}$\\ \hline \hline
$\varphi^{\5\ell,-1}_{-(\5\ell+1)}$	& $1$		& $+$		& $1$		& $-2$	& $2$		& $\frac{19}{10}$					& $\frac{14}{5}$	& ${\bf 45}$\\ \hline 
$\varphi^{\5\ell,-1}_{-(\5\ell+1)}$	& $1$		& $+$		& $3$		& $-4$	& $18$		& $\frac{17}{10}$					& $\frac{12}{5}$	& ${\bf 3}$\\ \hline \hline
\end{tabular}}}
\end{center}
There are $48$ non-aligned vectors and, hence, $108$ vectors in total.

\subsubsection*{Counting in $(2,2,8,3; 3)$}

\begin{center}
{\renewcommand{\arraystretch}{1.5}
\renewcommand{\tabcolsep}{0.5cm}
{\small \begin{tabular}{||c|c|c|c|c||c||} \hline \hline
\multicolumn6{||c||}{\bf 16 - aligned generations} \\ \hline \hline
$\sigma'$ 	& $\5\ell$	& $l$	& $\5q'_{\CP}$ 	& $q'_{\CP}$ 	& $N_{\bf 16}^A$\\ \hline \hline
$+$		& $0$		& $0$	& $\frac75$	& $\frac75$	& ${\bf13}$\\ \hline
$+$		& $2$		& $0$	& $\frac95$	& $\frac95$	& ${\bf7}$\\ \hline 
$+$		& $0$		& $2$	& $\frac75$	& $\frac75$	& ${\bf3}$\\ \hline
$+$		& $2$		& $2$	& $\frac95$	& $\frac95$	& ${\bf1}$\\ \hline \hline
 \end{tabular}}}
\end{center}
Hence, the number of aligned generations is $24$.
\begin{center}
 {\renewcommand{\arraystretch}{1.5}
\renewcommand{\tabcolsep}{0.5cm}
{\small \begin{tabular}{||c|c|c|c|c|c|c||c||} \hline \hline
\multicolumn8{||c||}{\bf 16 - non-aligned generations}\\ \hline \hline
Fermat					& $K\mod4$	& $\sigma'$ 	& $\5\ell$	& $\nu \mod 4K$	& $h'_{R}$ 	& $q'_{R}$ 	& $N_{\bf 16}^{NA}$\\ \hline \hline
$\varphi^{\5\ell,1}_{\5\ell+1}$		& $1$		& $-$		& $3$		& $16$		& $\frac12$						& $1$	& ${\bf 1}$\\ \hline 
$\varphi^{\5\ell,-1}_{-(\5\ell+1)}$	& $1$		& $+$		& $0$		& $24$		& $\frac12$							& $1$	& ${\bf 1}$\\ \hline  
$\varphi^{\5\ell,-1}_{-(\5\ell+1)}$	& $1$		& $+$		& $2$		& $20$		& $\frac{13}{10}$					& $\frac35$	& ${\bf 1}$\\ \hline 
$\varphi^{\5\ell,-1}_{-(\5\ell+1)}$	& $1$		& $+$		& $2$		& $40$		& $\frac{13}{10}$					& $\frac35$	& ${\bf 7}$\\ \hline \hline
 \end{tabular}}}
\end{center}
The number of non-aligned generations is $10$ which, together with the $24$ of the aligned generations gives a total of $34$ generations.
\begin{center}
{\renewcommand{\arraystretch}{1.5}
\renewcommand{\tabcolsep}{0.5cm}
{\small \begin{tabular}{||c|c|c|c|c||c||} \hline \hline
\multicolumn6{||c||}{\bf $\5{16}$ - aligned antigenerations}\\ \hline \hline
$\sigma'$ 	& $\5\ell$	& $l$	& $\5q'_{\CP}$ 	& $q'_{\CP}$ 	& $N_{\bf 16}^A$\\ \hline \hline
$-$		& $1$		& $5$	& $\frac35$	& $\frac85$	& ${\bf1}$\\ \hline 
$-$		& $3$		& $5$	& $1$		& $2$		& ${\bf1}$\\ \hline
$+$		& $0$		& $6$	& $1$		& $\frac75$	& ${\bf1}$\\ \hline 
$+$		& $0$		& $11$	& $1$		& $\frac75$	& ${\bf1}$\\ \hline 
$+$		& $0$		& $16$	& $1$		& $\frac75$	& ${\bf3}$\\ \hline
$+$		& $2$		& $13$	& $\frac35$	& $\frac95$	& ${\bf1}$\\ \hline  \hline
 \end{tabular}}}
\end{center}
Hence, there are $8$ aligned antigenerations.
\begin{center}
 {\renewcommand{\arraystretch}{1.5}
\renewcommand{\tabcolsep}{0.5cm}
{\small \begin{tabular}{||c|c|c|c|c|c|c||c||} \hline \hline
\multicolumn8{||c||}{\bf $\5{16}$ - non-aligned antigenerations}\\ \hline \hline
Fermat					& $K\mod4$	& $\sigma'$ 	& $\5\ell$	& $\nu \mod 4K$	& $h'_{R}$ 		& $q'_{R}$ 	& $N_{\bf 16}^{NA}$\\ \hline \hline
$\varphi^{\5\ell,1}_{\5\ell+1}$		& $1$		& $+$		& $2$		& $16$		& $\frac{9}{10}$	& $\frac95$	& ${\bf 1}$\\ \hline
$\varphi^{-\5\ell,1}_{-(\5\ell+1)}$	& $1$		& $-$		& $1$		& $32$		& $\frac{9}{10}$	& $\frac95$	& ${\bf 1}$\\ \hline \hline
 \end{tabular}}}
\end{center}
The number of non-aligned antigenerations is $2$. In total there are, hence, $10$ antigenerations.
\begin{center}
{\renewcommand{\arraystretch}{1.5}
\renewcommand{\tabcolsep}{0.5cm}
{\small \begin{tabular}{||c|c|c|c|c|c||c||} \hline \hline
\multicolumn7{||c||}{\bf 10 - aligned vectors}\\ \hline \hline
$q_\intern$	& $\sigma'$ 	& $\5\ell$	& l	& $\5q'_{\CP}$ 	& $q'_{\CP}$ 	& $N_{\bf10}^A$\\ \hline \hline
2		& $+$		& $1$		& $0$	& $\frac85$	& $\frac85$	& ${\bf10}$\\ \hline
2		& $+$		& $1$		& $10$	& $\frac85$	& $\frac85$	& ${\bf2}$\\ \hline
2		& $+$		& $3$		& $0$	& $2$		& $2$		& ${\bf3}$\\ \hline  \hline
1		& $-$		& $0$		& $5$	& $\frac75$	& $\frac25$	& ${\bf1}$\\ \hline
1		& $-$		& $2$		& $5$	& $\frac95$	& $\frac45$	& ${\bf1}$\\ \hline
1		& $+$		& $1$		& $0$	& $\frac85$	& $\frac45$	& ${\bf1}$\\ \hline
1		& $+$		& $1$		& $0$	& $\frac85$	& $\frac45$	& ${\bf3}$\\ \hline
1		& $+$		& $1$		& $0$	& $\frac85$	& $\frac45$	& ${\bf1}$\\ \hline\hline
 \end{tabular}}}
\end{center}
Hence, there are $22$ aligned vectors.
\begin{center}
 {\renewcommand{\arraystretch}{1.5}
\renewcommand{\tabcolsep}{0.5cm}
{\small \begin{tabular}{||c|c|c|c|c|c|c|c||c||} \hline \hline
\multicolumn9{||c||}{\bf 10 - non-aligned vectors}\\ \hline \hline
Fermat					& $K\mod4$	& $\sigma'$ 	& $\5\ell$	& $m$	& $\nu \mod 4K$	& $h'_{R}$ 					& $q'_{R}$ 	& $N_{\bf 10}^{NA}$\\ \hline \hline
$\varphi^{\5\ell,-1}_{-(\5\ell+1)}$	& $1$		& $+$		& $1$		& $-2$	& $2$		& $\frac{19}{10}$				& $\frac{14}{5}$& ${\bf 10}$\\ \hline 
$\varphi^{\5\ell,-1}_{-(\5\ell+1)}$	& $1$		& $+$		& $1$		& $-2$	& $22$		& $\frac{19}{40}$				& $\frac{14}{5}$& ${\bf 2}$\\ \hline
$\varphi^{\5\ell,-1}_m$			& $1$		& $-$		& $2$		& $-1$	& $32$		& $\frac{11}{10}$				& $2$		& ${\bf 1}$\\ \hline 
$\varphi^{\5\ell,1}_m$			& $1$		& $+$		& $1$		& $-6,+4$	& $38$		& $\frac{11}{10}$				& $2$	& ${\bf 1}$\\ \hline 
$\varphi^{\5\ell,1}_m$			& $1$		& $+$		& $1$		& $-6,+4$	& $18$		& $\frac{11}{10}$				& $2$	& ${\bf 1}$\\ \hline 
$\varphi^{\5\ell,1}_m$			& $1$		& $+$		& $1$		& $-6,+4$	& $8$		& $\frac{11}{10}$				& $2$		& ${\bf 3}$\\ \hline \hline
\end{tabular}}}
\end{center}
There are $48$ non-aligned vectors which, together with the $60$ aligned vectors, give a total of $108$ vectors.

\subsection{Non-Fermat-type examples}

\subsubsection*{Counting in $\mathbb{P}_{1,2,2,3,2}[10]$}

This model has $K=5$ and can therefore be used for checking the case where $K\equiv 1\mod4$.
The conjecture predicts equivalence with the DK model 
\be
V_{1,2,2,3,2}[10]\longrightarrow\mathbb{P}_{1,2,2,3,4,4}[8,8].
\ee
Its $\chi$-genus can be computed by applying (\ref{eq:LG-genus}) and we obtain
\ba
\begin{array}{||c|c||}
\hline  
\hline
 \alpha  & \chi _{\alpha } \\
\hline 
& \\
 0 & -t^4-55 t^3+55 t+1 \\
 1 & t^4 \\
 2 & -2 t^2 \\
 3 & t^2+t \\
 4 & -t \\
 5 & 5 t-5 t^3 \\
 6 & t^3 \\
 7 & -t^3-t^2 \\
 8 & 2 t^2 \\
 9 & -1 \\
 \hline
 \hline
\end{array}
\ea
Summing up the positive coefficients of monomials in $t$ and $t^3$, respectively, we get 61 generations and 1 antigeneration for the DK model.
Using our counting method we can compare this result with that on the CFT side.  
The relevant data for admissible terms in the EPP of $\C'$ for counting aligned generations is listed in the table below.
\begin{center}
{\renewcommand{\arraystretch}{1.5}
\renewcommand{\tabcolsep}{0.5cm}
{\small \begin{tabular}{||c|c|c|c|c||c||} \hline \hline
\multicolumn6{||c||}{\bf 16 - aligned generations}\\ \hline \hline
$\sigma'$ 	& $\5\ell$	& $l$	& $\5q'_{\CP}$ 	& $q'_{\CP}$ 	& $N_{\bf 16}^A$\\ \hline \hline
$+$		& $0$		& $0$	& $\frac75$	& $\frac75$	& ${\bf27}$\\ \hline
$+$		& $2$		& $0$	& $\frac95$	& $\frac95$	& ${\bf14}$\\ \hline 
$+$		& $0$		& $5$	& $\frac75$	& $\frac75$	& ${\bf3}$\\ \hline 
$+$		& $2$		& $5$	& $\frac95$	& $\frac95$	& ${\bf1}$\\ \hline \hline
 \end{tabular}}}
\end{center}
Hence, the number of aligned generations is $45$. The necessary information in order to count non-aligned generations in the {complementary Poincar\'{e}} polynomial is given by
\begin{center}
 {\renewcommand{\arraystretch}{1.5}
\renewcommand{\tabcolsep}{0.5cm}
{\small \begin{tabular}{||c|c|c|c|c|c|c||c||} \hline \hline
\multicolumn8{||c||}{\bf 16 - non-aligned generations}\\ \hline \hline
Fermat					& $K\mod4$	& $\sigma'$ 	& $\5\ell$	& $\nu \mod 4K$	& $h'_{R}$ 	& $q'_{R}$ 	& $N_{\bf 16}^{NA}$\\ \hline \hline
$\varphi^{\5\ell,-1}_{-(\5\ell+1)}$	& $1$		& $+$		& $0$		& $14$		& $\frac{1}{2}$					& $1$	& ${\bf 1}$\\ \hline
$\varphi^{\5\ell,-1}_{-(\5\ell+1)}$	& $1$		& $+$		& $2$		& $10$		& $\frac{3}{10}$					& $\frac35$	& ${\bf 1}$\\ \hline
$\varphi^{\5\ell,-1}_{-(\5\ell+1)}$	& $1$		& $+$		& $2$		& $0$		& $\frac{3}{10}$					& $\frac35$	& ${\bf 14}$\\ \hline \hline
 \end{tabular}}}
\end{center}
There are $16$ non-aligned vectors. In total thera are, hence, $61$ vectors which agrees with the prediction from the DK model. In order to count aligned antigenerations we need the following data.
\begin{center}
{\renewcommand{\arraystretch}{1.5}
\renewcommand{\tabcolsep}{0.5cm}
{\small \begin{tabular}{||c|c|c|c|c||c||} \hline \hline
\multicolumn6{||c||}{\bf $\5{16}$ - aligned antigenerations}\\ \hline \hline
$\sigma'$ 	& $\5\ell$	& $l$	& $\5q'_{\CP}$ 	& $q'_{\CP}$ 	& $N_{\bf \ovl{16}}^A$\\ \hline \hline

$+$		& $0$		& $6$	& $1$	& $\frac75$	& ${\bf1}$\\ \hline \hline
 \end{tabular}}}
\end{center}
Since there are no non-aligned antigenerations in this model there is in total only 1 antigeneration. This agrees with the prediction of the DK model.
Moreover, we predict the following data for aligned vectors.
\begin{center}
{\renewcommand{\arraystretch}{1.5}
\renewcommand{\tabcolsep}{0.5cm}
{\small \begin{tabular}{||c|c|c|c|c|c||c||} \hline \hline
\multicolumn7{||c||}{\bf 10 - aligned vectors}\\ \hline \hline
$q_\intern$	& $\sigma'$ 	& $\5\ell$	& l	& $\5q'_{\CP}$ 	& $q'_{\CP}$ 	& $N_{\bf10}^A$\\ \hline \hline
2		& $+$		& $1$		& $0$	& $\frac85$	& $\frac85$	& ${\bf21}$\\ \hline
2		& $+$		& $3$		& $0$	& $2$		& $2$		& ${\bf7}$\\ \hline
2		& $+$		& $1$		& $5$	& $\frac85$	& $\frac85$	& ${\bf2}$\\ \hline  \hline
1		& $+$		& $1$		& $2$	& $\frac85$	& $\frac45$	& ${\bf1}$\\ \hline
1		& $+$		& $1$		& $7$	& $\frac85$	& $\frac45$	& ${\bf1}$\\ \hline\hline
 \end{tabular}}}
\end{center}
Hence, there are $32$ aligned vectors.
\begin{center}
 {\renewcommand{\arraystretch}{1.5}
\renewcommand{\tabcolsep}{0.5cm}
{\small \begin{tabular}{||c|c|c|c|c|c|c|c||c||} \hline \hline
\multicolumn9{||c||}{\bf 10 - non-aligned vectors}\\ \hline \hline
Fermat					& $K\mod4$	& $\sigma'$ 	& $\5\ell$	& $m$	& $\nu \mod 4K$	& $h'_{R}$ 					& $q'_{R}$ 	& $N_{\bf 10}^{NA}$\\ \hline \hline
$\varphi^{\5\ell,1}_{(\5\ell+1)}$	& $1$		& $-$		& $0$		& $1$	& $2$		& $\frac{17}{10}$				& $\frac{12}{5}$& ${\bf 1}$\\ \hline 
$\varphi^{\5\ell,-1}_{-(\5\ell+1)}$	& $1$		& $+$		& $1$		& $-2$	& $22$		& $\frac{19}{40}$				& $\frac{14}{5}$& ${\bf 21}$\\ \hline
$\varphi^{\5\ell,-1}_{-(\5\ell+1)}$	& $1$		& $+$		& $1$		& $-2$	& $32$		& $\frac{11}{10}$				& $2$		& ${\bf 2}$\\ \hline 
$\varphi^{\5\ell,1}_m$			& $1$		& $+$		& $1$		& $-6,+4$& $18$		& $\frac{11}{10}$				& $2$		& ${\bf 1}$\\ \hline 
$\varphi^{\5\ell,1}_m$			& $1$		& $+$		& $1$		& $-6,+4$& $8$		& $\frac{11}{10}$				& $2$		& ${\bf 1}$\\ \hline \hline
\end{tabular}}}	
\end{center}
There are $26$ non-aligned vectors. We, therefore, predict a total number of $58$ vectors.

\subsubsection*{Counting in $\mathbb{P}_{1,1,2,2,1}[7]$}

This model has  $K=7$ and is, hence, a check for the case $K \equiv 3 \mod 4$.
The conjecture predicts equivalence with the DK model 
\be
V_{1,1,2,2,1}[7]\longrightarrow\mathbb{P}_{1,1,2,2,2,3}[6,5].
\ee
Its $\chi$-genus can be computed by applying (\ref{eq:LG-genus}) and we obtain
\ba
\begin{array}{||c|c||}
\hline  
\hline
 \alpha  & \chi _{\alpha } \\
\hline 
& \\
0 & -t^4-66 t^3+66 t+1 \\
 1 & t^4 \\
 2 & -t^2 \\
 3 & 3 t^2-3 t \\
 4 & 3 t^3-3 t^2 \\
 5 & t^2 \\
 6 & -1\\
 \hline
 \hline
\end{array}
\ea
Summing up the positive coefficients of monomials in $t$ and $t^3$, respectively, we get 66 generations and 3 antigenerations for the DK model.
On the CFT side we use the data in the tables below to count aligned generations and antigenerations.
\begin{center}
{\renewcommand{\arraystretch}{1.5}
\renewcommand{\tabcolsep}{0.5cm}
{\small \begin{tabular}{||c|c|c|c|c||c||} \hline \hline
\multicolumn6{||c||}{\bf 16 - aligned generations}\\ \hline \hline
$\sigma'$ 	& $\5\ell$	& $l$	& $\5q'_{\CP}$ 		& $q'_{\CP}$ 	& $N_{\bf 16}^A$\\ \hline \hline
$+$		& $0$		& $0$	& $\frac97$		& $\frac97$	& ${\bf36}$\\ \hline
$+$		& $2$		& $0$	& $\frac{11}{7}$	& $\frac{11}{7}$& ${\bf18}$\\ \hline 
$+$		& $4$		& $0$	& $\frac{13}{7}$	& $\frac{13}{7}$& ${\bf8}$\\ \hline \hline
 \end{tabular}}}
\end{center}
Hence, the number of aligned generations is $52$. 
\begin{center}
 {\renewcommand{\arraystretch}{1.5}
\renewcommand{\tabcolsep}{0.5cm}
{\small \begin{tabular}{||c|c|c|c|c|c|c||c||} \hline \hline
\multicolumn8{||c||}{\bf 16 - non-aligned generations}\\ \hline \hline
Fermat					& $K\mod4$	& $\sigma'$ 	& $\5\ell$	& $\nu \mod 4K$	& $h'_{R}$ 	& $q'_{R}$ 	& $N_{\bf 16}^{NA}$\\ \hline \hline
$\varphi^{\5\ell,-1}_{-(\5\ell+1)}$	& $3$		& $+$		& $3$		& $14$		& $\frac{2}{7}$					& $\frac47$	& ${\bf 14}$\\ \hline \hline
 \end{tabular}}}
\end{center}
There are $14$ non-aligned vectors. Hence, we get a total number of $66$ vectors which agrees with the prediction from the DK model. There are no aligned antigenerations in this model. In order to count non-aligned antigenerations we need the following data.
\begin{center}
 {\renewcommand{\arraystretch}{1.5}
\renewcommand{\tabcolsep}{0.5cm}
{\small \begin{tabular}{||c|c|c|c|c|c|c||c||} \hline \hline
\multicolumn8{||c||}{\bf $\5{16}$ - non-aligned antigenerations}\\ \hline \hline
Fermat					& $K\mod4$	& $\sigma'$ 	& $\5\ell$	& $\nu \mod 4K$	& $h'_{R}$ 	& $q'_{R}$ 	& $N_{\bf 16}^{NA}$\\ \hline \hline
$\varphi^{\5\ell,1}_{\5\ell+1}$		& $3$		& $+$		& $1$		& $8$		& $\frac{5}{7}$		& $\frac{10}{7}$& ${\bf 2}$\\ \hline 
$\varphi^{\5\ell,1}_{\5\ell+1}$		& $3$		& $+$		& $3$		& $8$		& $\frac{6}{7}$		& $\frac{12}{7}$& ${\bf 1}$\\ \hline \hline
 \end{tabular}}}
\end{center}
There are in total 3 antigenerations which is in agreement with the prediction of the DK model.

\section{Conclusions and perspectives}

In this paper, we have investigated the non-rational generalization \cite{bsw96}
of the CFT/geometry connection proposed 
 for $(0,2)$ heterotic compactifications in \cite{kr96,wkm}. 
To this aim, we first reformulated the construction of Blumenhagen et al.
\cite{bl95,bl96} in terms of simple current 
modular invariants identified with orbifolds with discrete torsion \cite{ks93,wkm}. In this language
the breaking of $E_6$ to the GUT gauge group $SO(10)$ is achieved thanks to the discrete torsions
associated with a simple current $J_b$ spoiling the algebra extension in the gauge sector
and corresponding to a $\IZ_2$ orbifold. 

We have proposed a simple counting algorithm for charged massless states. Counting in untwisted sectors
goes as in $(2,2)$ compactifications and can be reduced to the sector of BPS states. Instead, even for 
non-gauge-singlet states, 
the spectrum in $J_b$-twisted sectors gets contributions from 
non-BPS states that we analyzed in detail.

The counting algorithm can be used to compare the CFT side with the Distler-Kachru models appearing on the
geometry side. These are characterized by a rank 4 vector bundle $E$ on a Calabi-Yau 
manifold $X$ whose data are constrained by the anomaly matching condition 
$c_2(E)=c_2(X)$ and make sense also for certain non-rational {\em internal} superconformal theories
 like Landau-Ginzburg models and orbifolds thereof.
 
 While we focus on the SO(10) case, the generalization to $E_4=SU(5)$ and
$E_3=SU(3)\times SU(2)$ gauge groups is straightforward, at least on
the CFT side \cite{bl95}. Since
a minimal model factor is required in each reduction step, the number of these classes of 
models becomes slim in RCFT,  but the generalization to Landau-Ginzburg orbifolds
should partially make up for this and hopefully create some room
for interesting phenomenology. 

Besides, an important 
additional topic which could be explored with the methods of this paper is the singlet spectrum
which is interesting for
the study of deformations, in particular on the geometry side and
in combination with mirror symmetry for (0,2) models that are
not deformations of the tangent bundle \cite{Kreuzer:2010ph,Melnikov:2010sa,Aspinwall:2010ve}.

\bigskip
\noindent{\bf Acknowledgments}

\noindent
M.B. thanks J.~Distler and R.~Blumenhagen for kind discussions. A.P. wants to thank S.~Kachru for discussions and  the Kavli Institute for Theoretical Physics (KITP) Santa Barbara, where part of this work has been carried out, for their kind hospitality during the workshop ``Strings at the LHC and in the early Universe``. A.P. was partially supported by the Austrian Marshall Plan Foundation. M.K. and A.P. acknowledge support from the Austrian Research Funds FWF under grants number I192 and P21239.

In memoriam of Maximilian Kreuzer, who recently passed away,  M.B. and A.P  would like to thank him for his guidance in this work. With great sorrow we heartily remember Max for his great spirit and the enthusiasm and persistence he taught us in pursuing scientific ideas.

\newpage

\appendix
\section{Proof that $\mathscr{P}(x, 0, t) = P(t^{-1}, 1, x)$}
We want to compute
\be
\lim_{q\to 0} \left[q^{\frac{\widehat c}{2}\ell^{2}}\,t^{\widehat{c} \ell}\,\prod_{i=1}^{N}
  \frac{\vartheta_{1}(q, t^{1-\omega_{i}}\,q^{\ell(1-\omega_{i})})}
  {\vartheta_{1}(q, t^{\omega_{i}}\,q^{\ell\omega_{i}})}\right]
  = \lim_{q\to 0} \left[\prod_{i=1}^{N}
 q^{\frac{1-2\omega_{i}}{2}\ell^{2}}\,t^{(1-2\omega_{i}) \ell}\,
  \frac{\vartheta_{1}(q, t^{1-\omega_{i}}\,q^{\ell(1-\omega_{i})})}{\vartheta_{1}
  (q, t^{\omega_{i}}\,q^{\ell\omega_{i}})}\right].
\ee
Let us consider a specific superfield $\Phi_{i}$ and its contribution to the above limit. There are two possibilities.
If $\ell\omega_{i}\in\mathbb{N}$, the exponent of $q$ in $\vartheta_{1}(q, t^{1-\omega_{i}}\,
q^{\ell(1-\omega_{i})})$
is minimum for $n=-\ell(1-\omega_{i})$ and $n=-\ell(1-\omega_{i})+1$ with the same value. 
Similarly, the exponent of $q$ in $\vartheta_{1}(q, t^{\omega_{i}}\,q^{\ell\omega_{i}})$
is minimum for $n=-\ell\omega_{i}$ and $n=-\ell\omega_{i}+1$ with the same value. 
The contribution to 
$\mathscr{P}(x, 0, t^{-1})$ is a factor
\be
(-1)^{\ell}t^{-\frac{1-2\omega_{i}}{2}}\frac{1-t^{1-\omega_{i}}}{1-t^{\omega_{i}}},
\ee
If instead $\ell\omega_{i}\not\in\mathbb{N}$, let $\theta_{i}^{(\ell)} = \ell\omega_{i}-[\ell\omega_{i}]$. Let us assume
$0<\theta_{i}^{(\ell)}<1/2$ (a similar computation can be done in the case $1/2<\theta_{i}^{(\ell)}<1$).
The exponent of $q$ in $\vartheta_{1}(q, t^{1-\omega_{i}}\,q^{\ell(1-\omega_{i})})$
is minimum for $n=1-\ell+\ell\omega_{i}-\theta_{i}^{(\ell)}$. 
Similarly, the exponent of $q$ in $\vartheta_{1}(q, t^{\omega_{i}}\,q^{\ell\omega_{i}})$
is minimum for $n=\ell\omega_{i}-\theta_{i}^{(\ell)}$. 
The contribution to 
$\mathscr{P}(x, 0, t^{-1})$ is a factor
\be
(-1)^{\ell-1}t^{-\frac{1-2\omega_{i}}{2}}\,t^{\theta_{i}^{(\ell)}-\omega_{i}}.
\ee
Taking the product over the superfields  and writing 
\be
\prod_{\ell\omega_{i}\in\Z}(-1)^{\ell}t^{-\frac{1-2\omega_{i}}{2}}
\prod_{\ell\omega_{i}\not\in\Z}(-1)^{\ell-1}t^{-\frac{1-2\omega_{i}}{2}} = (-1)^{N-N_{tw}(\ell)}
t^{-\frac{\widehat{c}}{2}},
\ee
where $N_{tw}(\ell)$ is the number of twisted fields in the $\ell$-twisted sector, 
we recognize the EPP from \cite{kr95} evaluated at $\5t=1$.

\section{The quintic $3^{4}\otimes 3$: 
}

In order to derive the spectrum (80,0,74) of the (0,2) cousin of the
quintic \cite{bl95,bl96} we decompose the ``quintic Gepner model'' $3^5$
into $\mathcal C'=3^4$ and an additional Fermat factor $\Ph^5$, 
{\it i.e.} minimal model at level $k=3$, on which 
the Bonn-twist acts \cite{wkm}. 

We encode the charge degeneracies of the GSO-twisted but unprojected 
N=2 SCFT $\mathcal C'$, with alignment between $\C'$ and the Fermat factor, in its extended Poincar\'{e} polynomial \cite{kr95}:
For the untwisted sector we obtain the standard Poincar\'e polynomial (in the (c,c) ring)
\BEA	
	P(t,\bar t)&=&\textstyle\frac{(1-T^{4})^4}{(1-T)^4}=(1+T+T^2+T^3)^4 \nonumber
	~=~1+4T+10T^2+20T^3+
\\	&&+31 T^4+40 T^5+44 T^6+40 T^7
	+31T^8+20T^9+10T^{10}+4T^{11}+T^{12}
\label{Untwquintic}
\EEA
with $T=(t\bar t)^{1/5}$. In the twisted sectors only the ground states
contribute since there are no invariant fields. Hence the EPP continues
with the terms
\BE
	P(x,t^{5},\bar t^{5})=P(t^{5},\bar t^{5})+x\, \bar t\,^{12}
		+x^2 \,t^{4}\bar t\,^{8}+x^3 \,t^{8}\bar t\,^{4}
	+x^4 \,t^{12}+\ldots
\label{Twquintic}
\EE
and then ``periodically'' with $x^5P(t^{5},\bar t^{5})+x^6\, \bar t\,^{12}+\ldots$
\\For the (2,2) version of the quintic we would multiply with an additional
$1+T+T^2+T^3$ and obtain the famous $101=10+20+31+40$ from $P(t,\bar t)$, which is
the K\"ahler modulus from the $x^2$-term.
{In order to determine the number of aligned generations, antigenerations and vectors for the (0,2) cousin we read off the relevant data from the tables \ref{Agen},\ref{Aantigen} and \ref{Avec} to get $N^A_{\bf 16}=60$ from the $T^7$ and $T^9$ terms, $N^A_{\bf \5{16}}=0$ and $N^A_{\bf 10}=74$ from the $T^8$ and $T^{10}$ terms in \ref{Untwquintic} and from the $x^2t^4 \5t^8$ term in \ref{Twquintic}.}

The complementary Poincar\'{e} polynomial $\mathscr{P}(x, q, t)$ reads (up to ${\cal O}(q^{8/5})$ terms)
\begin{eqnarray}
\mathscr{P}(x, q, t^{5}) &=& \frac{1}{t^6}+\frac{4}{t^5}+\frac{10}{t^4}+\frac{20}{t^3}+\frac{31}{t^2}+\frac{40}{t}+44+40 t+31 t^2+{\bf 20 t^3}+10 t^4+4t^5+t^6 \nonumber \\
   &+&
   x\left[
  t^6+q^{1/5} \left(-4t^2+4t^7\right)+q^{2/5} \left(\frac{6}{t^2}-16t^3+10t^8\right)+q^{3/5}\left(-\frac{4}{t^6}+\frac{24}{t}- \right. \right. \nonumber \\ && \left. \left. \qquad -40t^4+20t^9\right)+q^{4/5}\left(60+\frac{1}{t^{10}}-\frac{16}{t^5}-76t^5+31t^{10}\right) + \cdots + \right. \nonumber \\  &&\left. \qquad +q^{8/5}\left(\frac{4}{t^{11}}-\frac{57}{t^6}+\frac{168}{t}-150t^4+4t^9+{\bf 31t^{14}}\right)
  \right] \nonumber\\ 
   &+&
   x^{2}\left[t^2+q^{2/5} \left(-\frac{4}{t^2}+4t^3\right)+q^{3/5}\left(4t-4t^6\right)+q^{4/5}\left(\frac{6}{t^6}-\frac{16}{t}+10t^4\right)+ \right. \nonumber \\  &&\left. \qquad +\cdots +q^{8/5}\left(\frac{1}{t^{14}}-\frac{16}{t^9}+\frac{20}{t^4}+28t-57t^6+24t^{11}\right)
   \right] \nonumber \\
   &+&
   x^{3}\left[
   \frac{1}{t^2}+q^{2/5} \left(\frac{4}{t^3}-4t^2\right)+q^{3/5}\left(-\frac{4}{t^6}+\frac{4}{t}\right)+q^{4/5}\left(\frac{10}{t^4}-16t+6t^6\right)+ \right. \nonumber \\&& \left. \qquad \cdots +q^{8/5}\left(\frac{24}{t^{11}}-\frac{57}{t^6}+\frac{28}{t}+20t^4-16t^9+t^{14}\right)
   \right] \nonumber \\
   &+&
   x^{4}\left[
  \frac{1}{t^6}+q^{1/5} \left(\frac{4}{t^7}-\frac{4}{t^2}\right)+q^{2/5} \left(\frac{10}{t^8}-\frac{16}{t^3}+6t^2\right)+q^{3/5}\left(\frac{20}{t^9}-\frac{40}{t^4}+ \right. \right. \nonumber \\ &&\left. \left. \qquad +24t-4t^6\right)+q^{4/5}\left(60+\frac{31}{t^{10}}-\frac{76}{t^5}-16t^5+{\bf t^{10}}\right)+\cdots+ \right. \nonumber \\ && \left. \qquad +q^{8/5}\left(\frac{31}{t^{14}}+\frac{4}{t^9}-\frac{150}{t^4}+168t-57t^6+4t^{11}\right)
   \right]  + \cdots, 
\label{Pquintic}
\end{eqnarray}
with the next terms being ``periodic'' in $x$.
{The number of non-aligned generations, antigenerations and vectors as read off from \ref{Pquintic} using the information of the tables \ref{NAgen}, \ref{NAantigen} and \ref{NAvec}. We get $N^{NA}_{\bf 16}=20$ from the $q^0 t^3$ term and $N^{NA}_{\bf 10}=32$ from the coefficients of $x q^{8/5}t^{14}$ and $x^4q^{4/5}t^{10}$.}

\del
\section{List of models with degree $\le 15$}
\label{app:models}

We list all models with degree $\le 15$. Notice that cases where $C'$ is not transverse cannot be treated.
\bigskip

\newcount\nIV	\newcount\nV	\newcount\dI	\newcount\dII  
\newcount\ChIII	\newcount\CUBE

\def\MMGM#1 #2 #3 #4 {\ifnum #2<6 \,\mathbf F \fi}
\def\WPH#1 #2 #3 #4 #5 #6 M:#7 V:#8 [#9]{ \nIV=0 \dI=0 \ChIII=0 \VR52
    	\global\advance\nIV by #6	\global\multiply\nIV by 2
    	\global\advance\dI by #1  	\global\advance\dI by -#6
	\nV=\dI				\global\divide\nV by 2
	\dII=\nIV			\global\advance\dII by \the\nV
        \CUBE=0	\global\advance\CUBE by #1	\global\multiply\CUBE by #1
	\global\multiply\CUBE by #1	\global\advance\ChIII by \the\CUBE
        \CUBE=0	\global\advance\CUBE by #6	\global\multiply\CUBE by #6
	\global\multiply\CUBE by #6	\global\advance\ChIII by -\the\CUBE
        \CUBE=\nIV			\global\multiply\CUBE by \the\nIV
	\global\multiply\CUBE by\the\nIV\global\advance\ChIII by \the\CUBE
        \CUBE=\nV			\global\multiply\CUBE by \the\nV
	\global\multiply\CUBE by\the\nV	\global\advance\ChIII by \the\CUBE
        \CUBE=\dI			\global\multiply\CUBE by \the\dI
	\global\multiply\CUBE by\the\dI	\global\advance\ChIII by -\the\CUBE
        \CUBE=\dII			\global\multiply\CUBE by \the\dII
	\global\multiply\CUBE by\the\dII\global\advance\ChIII by -\the\CUBE

	$\mathbb \IP_{#2,#3,#4,#5,#6}^{\MMGM#7 }[#1]_{#9}^{#8}$
	& $\IP_{#2,#3,#4,#5,\the\nIV,\the\nV}[\the\dI,\the\dII]_{
								}$}

\centerline{\footnotesize\noindent	\begin{tabular}{||c|c||}
\hline\hline
\WPH	5 1 1 1 1 1 M:126 5 N:6 5 V:1,101 [-200] 	\\\hline
\WPH	10 1 2 2 3 2 M:87 8 N:8 6 V:3,75 [-144]		\\\hline
\WPH	10 1 1 3 3 2 M:111 9 N:8 6 V:3,87 [-168]	\\\hline
\WPH	10 1 1 2 4 2 M:126 8 N:8 6 V:3,99 [-192]	\\\hline
\WPH	10 1 1 1 5 2 M:196 5 N:7 5 V:1,145 [-288]	\\\hline
\WPH	13 2 3 3 4 1 M:79 13 N:10 8 V:5,62 [-114]	\\\hline
\WPH	13 1 3 4 4 1 M:119 10 N:10 7 V:5,92 [-174]	\\\hline
	\hline\end{tabular}\begin{tabular}{c|c||}\hline\hline
\WPH	13 2 2 3 5 1 M:94 14 N:10 9 V:5,73 [-136]	\\\hline
\WPH	13 1 3 3 5 1 M:126 10 N:10 7 V:5,97 [-184]	\\\hline
\WPH	13 1 2 3 6 1 M:156 11 N:10 7 V:5,117 [-224]	\\\hline
\WPH	13 1 1 4 6 1 M:233 9 N:10 6 V:5,173 [-336]  \\\hline
\WPH	15 2 3 3 4 3 M:53 9 N:10 6 V:5,53 [-96]	\\\hline  
\WPH	15 1 3 4 4 3 M:71 9 N:10 6 V:9,54 [-90]	\\\hline
\WPH	15 2 2 3 5 3 M:56 9 N:8 6 V:7,43 [-72]		\\\hline
	\hline\end{tabular}\begin{tabular}{c|c||}\hline\hline
\WPH	15 1 3 3 5 3 M:81 5 N:9 5 V:3,75 [-144] 	\\\hline
\WPH	15 1 2 4 5 3 M:84 10 N:8 7 V:3,66 [-126]	\\\hline
\WPH	15 1 1 5 5 3 M:134 5 N:8 5 V:7,103 [-192]	\\\hline
\WPH	15 1 2 3 6 3 M:96 10 N:10 7 V:5,77 [-144]	\\\hline
\WPH	15 1 1 4 6 3 M:139 11 N:11 8 V:7,106 [-198]	\\\hline
\WPH	15 1 2 2 7 3 M:118 11 N:11 8 V:5,89 [-168]	\\\hline
\WPH	15 1 1 3 7 3 M:158 6 N:10 6 V:9,117 [-216]	\\\hline
\hline
\end{tabular}}
\bigskip
\centerline{Table I: Examples with $K\equiv 5$ modulo 4 where  F=``Fermat type''}

\bigskip
\bigskip
\centerline {\footnotesize\noindent	\begin{tabular}{||c|c||}
\hline\hline
\WPH	6 1 1 1 1 2 M:130 5 N:6 5 V:1,103 [-204]	\\\hline
\WPH	7 1 1 2 2 1 M:120 9 N:7 6 V:2,95 [-186]		\\\hline
\WPH	7 1 1 1 3 1 M:159 8 N:7 6 V:2,122 [-240]	\\\hline
\WPH	9 1 1 2 2 3 M:109 9 N:7 6 V:2,86 [-168]		\\\hline
\WPH	9 1 2 2 3 1 M:109 9 N:7 6 V:2,86 [-168]		\\\hline
\WPH	9 1 1 1 3 3 M:145 5 N:7 5 V:4,112 [-216]	\\\hline
\WPH	9 1 1 3 3 1 M:145 5 N:7 5 V:4,112 [-216]	\\\hline
\WPH	9 1 1 1 2 4 M:162 9 N:8 6 V:3,123 [-240]	\\\hline
\WPH	11 2 2 3 3 1 M:81 13 N:9 7 V:4,64 [-120]	\\\hline
\WPH	11 1 2 3 4 1 M:121 13 N:9 8 V:4,94 [-180]	\\\hline
\WPH	11 1 2 2 5 1 M:144 10 N:9 7 V:4,109 [-210]	\\\hline
\WPH	11 1 1 3 5 1 M:192 10 N:9 7 V:4,144 [-280]	\\\hline
\WPH	12 1 2 2 3 4 M:89 5 N:7 5 V:2,74 [-144]		\\\hline
	\hline\end{tabular}\begin{tabular}{c|c||}\hline\hline
\WPH	12 1 1 3 3 4 M:115 5 N:7 5 V:5,89 [-168]	\\\hline
\WPH	12 1 1 2 4 4 M:130 5 N:8 5 V:5,101 [-192]	\\\hline
\WPH	14 2 3 3 4 2 M:57 10 N:10 7 V:5,51 [-92]		\\\hline
\WPH	14 1 2 4 5 2 M:99 10 N:10 7 V:5,83 [-156]		\\\hline
\WPH	14 1 2 3 6 2 M:108 9 N:10 6 V:5,85 [-160]	\\\hline
\WPH	14 1 1 4 6 2 M:159 10 N:10 7 V:5,121 [-232]	\\\hline
\WPH	14 1 2 2 7 2 M:141 5 N:9 5 V:2,122 [-240]	\\\hline
\WPH	14 1 1 3 7 2 M:179 7 N:8 6 V:2,132 [-260]	\\\hline
\WPH	15 3 3 4 4 1 M:71 9 N:10 6 V:9,54 [-90]	\\\hline
\WPH	15 2 2 3 3 5 M:56 9 N:8 6 V:7,43 [-72]	\\\hline
\WPH	15 1 3 3 3 5 M:81 5 N:9 5 V:3,75 [-144]	\\\hline
\WPH	15 3 3 3 5 1 M:81 5 N:9 5 V:3,75 [-144]	\\\hline
\WPH	15 1 2 3 4 5 M:84 10 N:8 7 V:3,66 [-126]	\\\hline
	\hline\end{tabular}\begin{tabular}{c|c||}\hline\hline
\WPH	15 2 3 4 5 1 M:84 10 N:8 7 V:3,66 [-126]	\\\hline
\WPH	15 1 2 2 5 5 M:100 9 N:9 6 V:8,77 [-138]	\\\hline
\WPH	15 2 2 5 5 1 M:100 9 N:9 6 V:8,77 [-138]	\\\hline
\WPH	15 1 1 3 5 5 M:134 5 N:8 5 V:7,103 [-192]	\\\hline
\WPH	15 1 3 5 5 1 M:134 5 N:8 5 V:7,103 [-192]	\\\hline
\WPH	15 2 3 3 6 1 M:96 10 N:10 7 V:5,77 [-144]	\\\hline
\WPH	15 1 3 4 6 1 M:139 11 N:11 8 V:7,106 [-198]	\\\hline
\WPH	15 2 2 3 7 1 M:118 11 N:11 8 V:5,89 [-168]	\\\hline
\WPH	15 1 3 3 7 1 M:158 6 N:10 6 V:9,117 [-216]	\\\hline
\WPH	15 1 2 4 7 1 M:177 12 N:11 8 V:6,132 [-252]	\\\hline
\WPH	15 1 1 5 7 1 M:282 7 N:10 6 V:4,208 [-408]	\\\hline
\WPH	15 1 1 1 7 5 M:282 7 N:10 6 V:4,208 [-408]	\\\hline
&\VR24\\\hline\hline
\end{tabular}}
\bigskip

\centerline{Table II: Examples with $K\equiv 3$ modulo 4, where  F=``Fermat type''}
\enddel

\end{document}